\documentclass[aps,pre,showpacs]{revtex4}

\usepackage{amsmath,epsfig,amsfonts,graphicx,
}

\begin{document}
\title{Highly Nonlinear Solitary Waves in Heterogeneous Periodic Granular Media}
\author{Mason A. Porter$^{1}$,  Chiara Daraio$^{2,*}$, Ivan Szelengowicz$^{2}$,
Eric B. Herbold$^3$,  and P. G. Kevrekidis$^4$}
\affiliation{ 
$^1$Oxford Centre for Industrial and Applied Mathematics, Mathematical Institute, University of Oxford, OX1 3LB, United Kingdom \\
$^2$Graduate Aeronautical Laboratories (GALCIT) and Department of Applied Physics,  \\ 
California Institute of Technology, Pasadena, CA 91125, USA \\
$^3$Department of Mechanical and Aerospace Engineering, University of California at San Diego, La Jolla, California 92093-0411, USA \\
$^4$Department of Mathematics and Statistics, University of Massachusetts, Amherst MA 01003-4515, USA }

\begin{abstract}
We use experiments, numerical simulations, and theoretical analysis to investigate the propagation of highly nonlinear solitary waves in periodic arrangements of dimer (two-mass) and trimer (three-mass) cell structures in one-dimensional granular lattices.  To vary the composition of the fundamental periodic units in the granular chains, we utilize beads of different  materials (stainless steel, bronze, glass, nylon, polytetrafluoroethylene, and rubber). This selection allows us to tailor the response of the system based on the masses, Poisson ratios, and elastic moduli of the components.  For example, we examine dimer configurations with two types of heavy particles, two types of light particles, and alternating light and heavy particles.  Employing a model with Hertzian interactions between adjacent beads, we find very good agreement between experiments and numerical simulations.  We find equally good agreement between these results and a theoretical analysis of the model in the long-wavelength regime that we derive for heterogeneous environments (dimer chains) and general bead interactions.  Our analysis encompasses previously-studied examples as special cases and also provides key insights on the influence of heterogeneous lattices on the properties (width and propagation speed) of the nonlinear wave solutions of this system.
\end{abstract}


\maketitle


\section{Introduction}  

Ever since the Fermi-Pasta-Ulam problem was first investigated half a 
century ago, nonlinear oscillator chains have received a great deal of 
attention \cite{fpu55,focus,Ford,lich}.  Over the past few years, in particular, such chains of nonlinear oscillators have proven to be important in numerous areas of physics--including Bose-Einstein condensation in optical lattices in atomic physics \cite{morsch1,konotopmplb,fpubec}, coupled waveguide arrays and photorefractive crystals in nonlinear optics \cite{photon,efrem1}, and DNA double-strand dynamics in biophysics \cite{peyrard}.  

The role of ``heterogeneous" versus ``uniform" lattices and the interplay between nonlinearity and periodicity has been among the key themes in investigations of lattice chains \cite{focus2}.  Here we investigate this idea using one-dimensional (1D) lattices of granular materials, which consist of chains of interacting spherical particles that deform elastically when they collide.  The highly nonlinear dynamic response of such lattices has been the subject of considerable attention from the scientific community 
\cite{nesterenko1,coste97,nesterenko2,dar05,dar05b,dar06,hascoet00,hinch99,man01,hong02,hong05,job05,doney06,ver06,sok07,herb06}.  In contrast to the traditionally-studied nonlinear oscillator chains, these systems possess a ``double" nonlinearity arising from the nonlinear contact interaction between the particles and a zero tensile response. This results in an asymmetric potential, which has in turn led to the observation of several new phenomena.  In particular, some of the most interesting dynamic properties of such highly nonlinear systems appear when the materials are under precompression \cite{nesterenko1,coste97,nesterenko2}, at the interface between two different highly nonlinear systems \cite{dar05b,dar06,hong02,hong05}, or at the interface between linear and nonlinear structures \cite{job05}.  This completely new type of wave dynamics opens the door for exciting new fundamental physical phenomena and has the potential to ultimately yield numerous new devices and materials.  Because granular lattices can be created from numerous material types and sizes, their properties are extremely tunable \cite{nesterenko1,nesterenko2,coste97}.  
Furthermore, the addition of a static precompressive force can substantially vary the response of such systems and allows the selection of the wave's regime of propagation between highly nonlinear and weakly nonlinear dynamics \cite{hong01b}.  Such tunability is valuable not only for studies of the basic physics of granular lattices but also in potential engineering applications.  Proposed uses include shock and energy absorbing layers \cite{dar06,hong05,doney06,ver06}, sound focusing devices (tunable acoustic lenses and delay lines), sound absorption layers, sound scramblers \cite{dar05,dar05b,herb06}, and more.

In the present work, we focus on one of the fundamental physical properties 
of such chains.  Namely, because they are highly nonlinear, granular lattices admit a novel type of wave solution whose qualitative properties differ markedly from those in weakly nonlinear systems.  Indeed, it was the experimental realization of such waves and the theory developed for uniform lattice systems of this type that has motivated the current broad interest in such 
settings \cite{nesterenko1}.  The remarkable property of these waves is that they essentially possess a support that consists of of just a few lattice sites \cite{nesterenko1}, providing perhaps the closest experimentally tractable application of the notion of ``compactons'' \cite{rosenau1} and establishing the potential for the design and creation of systems with unprecedented properties. 

Our concern in this paper, which provides a detailed and expanded 
description of work we reported in a recent letter \cite{dimer}, is to extend the established theory for uniform granular lattices to nonuniform ones.  One way to do this is to study the effects of defects, such as inhomogeneities, particles with different masses, and so on.  This has led to the observation of interesting physical responses such as fragmentation, anomalous reflections, and energy trapping \cite{hascoet00,ver06,hinch99,man01,hong02,hong05,job05,dar06,doney06}.  In the present paper, on the other hand, we examine the prototypical heterogeneities of granular lattice ``dimers'' and ``trimers,'' consisting of chains of multiparticle cells composed, respectively, of two and three different types of beads.  Such heterogeneities arise in a diverse array of physical settings, including ferroelectric perovskites \cite{dimer81,ferro} and polymers \cite{polymer}, optical waveguides \cite{sukh02,moran04}, and cantilever arrays \cite{sievers}.  
In our granular setting, we use a variety of soft, hard, heavy, and light materials to investigate the effects of different structural properties in the fundamental components of such systems.  We also vary the number of beads of a given type in each cell in order to examine the effects of different unit cell sizes (i.e., different periodicities).

More specifically, we investigate solitary wave propagation using experiments, numerical simulations, and theoretical analysis.  We report very good agreement between experiments and numerics.  For the case of dimer chains, we also construct a long-wavelength approximation to the nonlinear lattice model to obtain a quasi-continuum nonlinear partial differential equation (PDE) that provides an averaged 
description of the system.  We obtain analytical expressions for wave solutions of this equation and find very good qualitative agreement between the width and propagation speed of these solutions with those 
obtained from experiments and numerical simulations.  

The rest of this paper is organized as follows.  First, we present our experimental and numerical setups.  We then consider chains of dimers and discuss our experimental and numerical results.  We subsequently use a long-wavelength approximation to derive a nonlinear PDE describing the dimer setup in the case of general power-law interactions between materials and construct analytical expressions for the propagation speed, width, and functional form of its solitary wave solutions.  We then compare this continuum theory to our experiments and numerical simulations of the discrete system.  Finally, we discuss our experiments and numerical simulations for chains of trimers and summarize our results.


\section{Experimental Setup}

\begin{figure}[tbp]
\centering \includegraphics[width=150mm]{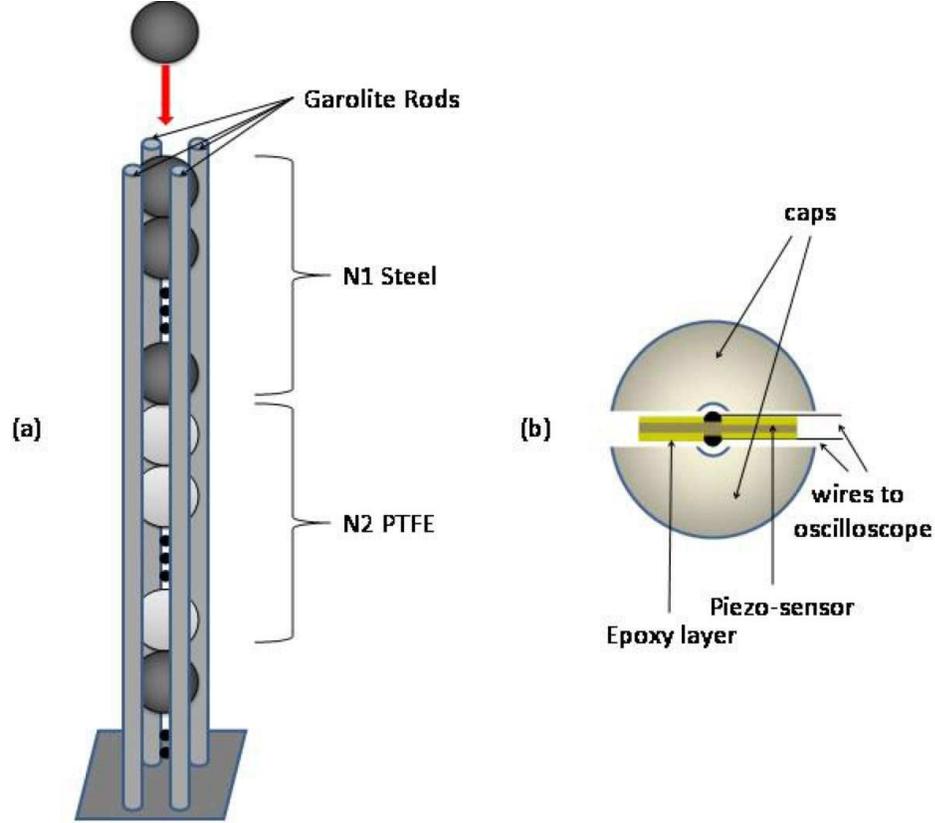}
\caption{(Color online) (a) Experimental setup for a dimer chain consisting of a periodic array of  cells with $N_1$ consecutive beads of one material (e.g., stainless steel) and $N_2$ consecutive beads of another material (e.g., PTFE).  (b) Schematic diagram of the composition of the sensors placed in the chain.
}
\label{setup}
\end{figure}

The experimental dimer and trimer chains were composed of vertically aligned beads in a delrin guide that contained slots for sensor connections or in a guide composed of four vertical garolite rods arranged in a square lattice [see Fig.~\ref{setup}(a)].  Each ``$N_1:N_2$ dimer" consisted of a variable number $N_1 \in \{1,\dots, 7\}$ of one type of bead alternating with $N_2 \in \{1,\dots, 7\}$ of a second type of bead in a periodic sequence.  The ``$N_1:N_2:N_3$ trimers" we studied are defined analogously.  The predominant class of configurations we considered included $N_1$ high-modulus, large mass stainless steel beads (non-magnetic, 316 type) and $N_2$ low-modulus, small mass polytetrafluoroethylene (PTFE) or Neoprene rubber elastomer  beads.  We also examined steel:bronze, PTFE:glass, and PTFE:nylon dimers and 1:1:1 trimers of steel:bronze:PTFE, steel:glass:nylon, and steel:PTFE:rubber.  The diameter of all spheres was 4.76 mm. 

For each experiment, we connected four calibrated piezosensors to a Tektronix oscilloscope (TDK2024) to detect force-time curves.  Three of the piezo-sensors were embedded inside particles in the chain and a fourth one was positioned at the bottom (i.e., at the wall).  To fabricate the sensors, beads selected from the various materials were cut in two halves and slots for sensors and wires were carved within them. Vertically poled lead zirconate titanate piezo elements (square plates with 0.5 mm thicknesses and 3 mm sides), supplied by Piezo Systems, Inc. (RC=$10^3$ microseconds), were soldered to custom micro-miniature wires and glued between the two bead halves [see Fig.~\ref{setup}(b)].  We calibrated the setup using conservation of momentum.  Waves were generated in the chains by dropping a striker from various heights.  In most of our experiments, the striker consisted of a stainless steel bead; for some configurations, we also used a PTFE bead, a glass bead, and a PTFE:glass dimer.  The material properties for the various beads are shown in Table \ref{parameters}.


\begin{table}
\centerline{
\begin{tabular}{|c|c|c|c|} \hline
Material & mass & $E$ & $\nu$ \\ \hline 
Steel & 0.45 g & 193 GPa & 0.3 \\ \hline
PTFE & 0.123 g & 1.46 GPa & 0.46 \\ \hline
Rubber & 0.08 g & 30 MPa & 0.49 \\ \hline
Bronze & 0.48 g & 76 GPa & 0.414 \\ \hline
Glass & 0.137 g & 62 GPa & 0.2 \\ \hline
Nylon & 0.0612 g & 3.55 GPa & 0.4 \\ \hline
\end{tabular}}
\caption{Material properties (mass, elastic modulus $E$, and Poisson ratio $\nu$) for stainless steel \cite{metals,316}, PTFE \cite{dar05,dupont,carter95}, rubber (McMaster-Carr), bronze \cite{bronze,bronzepoisson}, glass \cite{glass}, and nylon \cite{coste99}.  The value of the dynamic elastic modulus of the rubber beads was extrapolated from the experimental data.  
}
\label{parameters}
\end{table}

\section{Numerical simulations}  

We model a chain of $n$ spherical beads as a 1D lattice with Hertzian interactions between beads \cite{nesterenko1}:
\begin{align}
	\ddot{y}_j &= \frac{A_{j-1,j}}{m_j}\delta_{j}^{3/2} - \frac{A_{j,j+1}}{m_j}\delta_{j+1}^{3/2} + g \,, \notag \\
	A_{j,j+1} &= \frac{4E_{j}E_{j+1}\left(\frac{R_jR_{j+1}}{R_j + R_{j+1}}\right)^{1/2}}{3\left[E_{j+1}\left(1-\nu_{j}^2\right) + E_j\left(1-\nu_{j+1}^2\right)\right]}\,,	 \label{motion}
\end{align}
where $j \in \{1,\cdots,n\}$, $y_j$ is the coordinate of the center of the $j$th particle, $\delta_j \equiv \mbox{max}\{y_{j-1} - y_{j},0\}$ for $j \in \{2,\dots,n\}$, $\delta_1 \equiv 0$, $\delta_{n+1} \equiv \mbox{max}\{y_{n},0\}$, $g$ is the gravitational acceleration, $E_j$ is the Young's (elastic) modulus of the $j$th bead, $\nu_j$ is its Poisson ratio, $m_j$ is its mass, and $R_j$ is its radius.  The particle $j = 0$ represents the steel striker, and the $(n+1)$st particle represents the wall (i.e., an infinite-radius particle that cannot be displaced).  Our numerical simulations incorporate the nonuniform gravitational preload due to the vertical orientation of the chains in experiments but do not take dissipation into account.

The initial velocity of the striker is determined by experiments and all other particles start at rest in their equilibrium positions, which are are determined by solving a statics problem ($\ddot{y}_j = 0$ for all $j$).  Each bead experiences a force from gravity (one can also include a constant precompression $f_j$, but that is zero for the problem we study).  One starts by considering the bottom of the chain with particle $n$ against the wall.  It experiences a force $F_1$ from the top that is the sum of all the forces acting through the center of particle $n$ and a force $F_2$ from the wall (particle $n+1$) acting through the nonlinear spring (the Hertzian interaction).  The expressions for $F_1$ and $F_2$,
\begin{align}
	F_1 &= \sum_{j = 1}^{n}m_jg \,, \notag \\
	F_2 &= m_{n}A_{n,n+1}(u_{n} - u_{n+1})^{3/2}\,, \label{staticsetup}
\end{align}	
must be equal at equilibrium.  The displacement of the wall $u_{n+1} = 0$ is known, so Eq.~(\ref{staticsetup}) can be solved for the equilibrium displacement of particle $n$ in terms of known quantities:
\begin{equation}
	u_{n} = \left(\frac{\sum_{j = 1}^nm_jg}{A_{n,n+1}}\right)^{2/3} + u_{n+1}\,. \label{equilib}
\end{equation}
One then repeats this analysis one particle at a time (from the bottom of the chain to the top) to obtain the other equilibrium positions.

\section{Chains of Dimers}

Let us first discuss our results for chains of dimers.  While we focus our presentation on steel:PTFE dimers, we also investigated steel:rubber, steel:bronze, PTFE:glass, and PTFE:nylon dimers.  Additionally, our presentation includes discussions of 1:1, $N_1:1$, and $1:N_2$ dimers.  To compare numerical simulations with experiments, for which the sensors are inserted inside the beads rather than at the points of contact [see panel (b) of Fig.~\ref{setup}], we averaged the force between adjacent beads, $F_m=(F_j+F_{j+1})/2$, as discussed in detail in Ref.~\cite{dar05}.

Our numerical and experimental results greatly expand the previous work reported for chains of dimers composed of particles with similar elastic moduli and different masses \cite{nesterenko1}.  For example, we include investigations of dimers composed of particles with moduli of different orders of magnitude.  We also study dimers composed of different numbers of particles ($N_1:1$ and $1:N_2$ chains), thereby  varying the ``thickness" of the heterogeneous layers.

\subsection{1:1 Dimers}

\begin{figure}[tbp] 
\centering{\includegraphics[width=15.0cm]{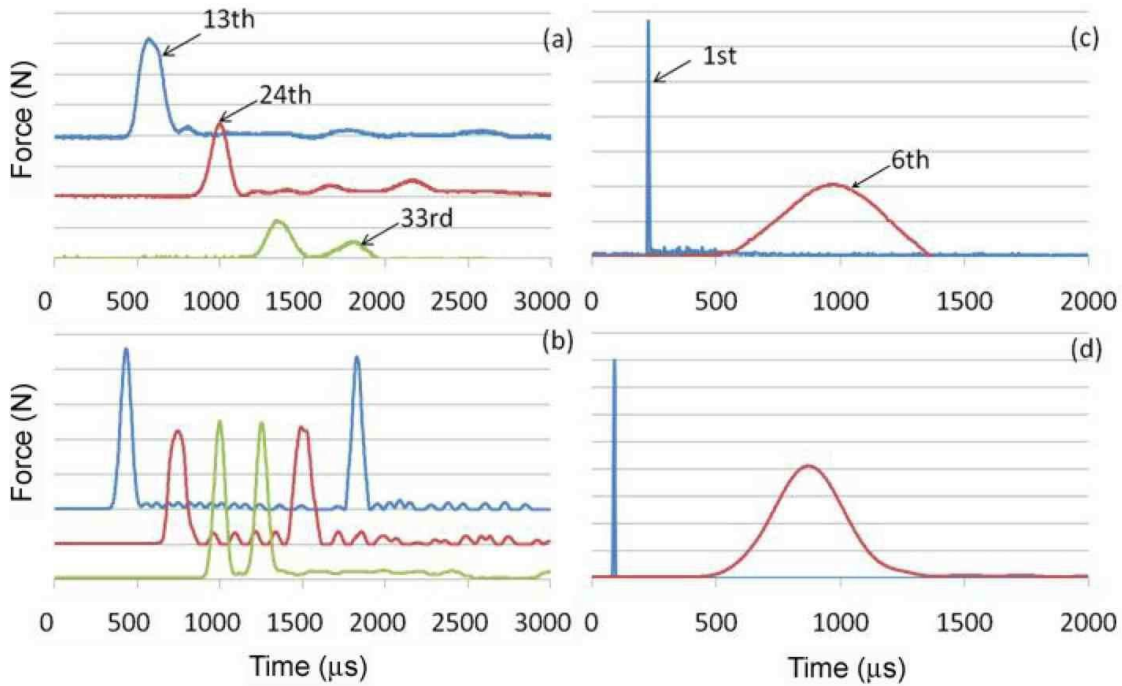}}
\caption{(Color online)  Force versus time response obtained from chains of dimers consisting of 1 stainless steel bead alternating with 1 PTFE (a,b) or 1 rubber (c,d) bead.  Panels (a,c) show experimental results, and (b,d) show the corresponding numerical data.  For both configurations, the initial velocity of the striker was 1.37 m/s on impact.  
The $y$-axis scale is 2 N per division in (a,b) and 20 N per division in (c,d).  The numbered arrows point to the corresponding particles in the chain.  In (a,b), the second curve (showing the results for particle 24) represents a PTFE bead and the other curves represent steel beads.  In (c,d), particle 1 is steel and particle 6 is rubber.  The force on the rubber particle is magnified by a factor of 20 for clarity.
}
\label{oneone}
\end{figure}

\begin{figure}[tbp]
\centering{\includegraphics[width=15.0cm]{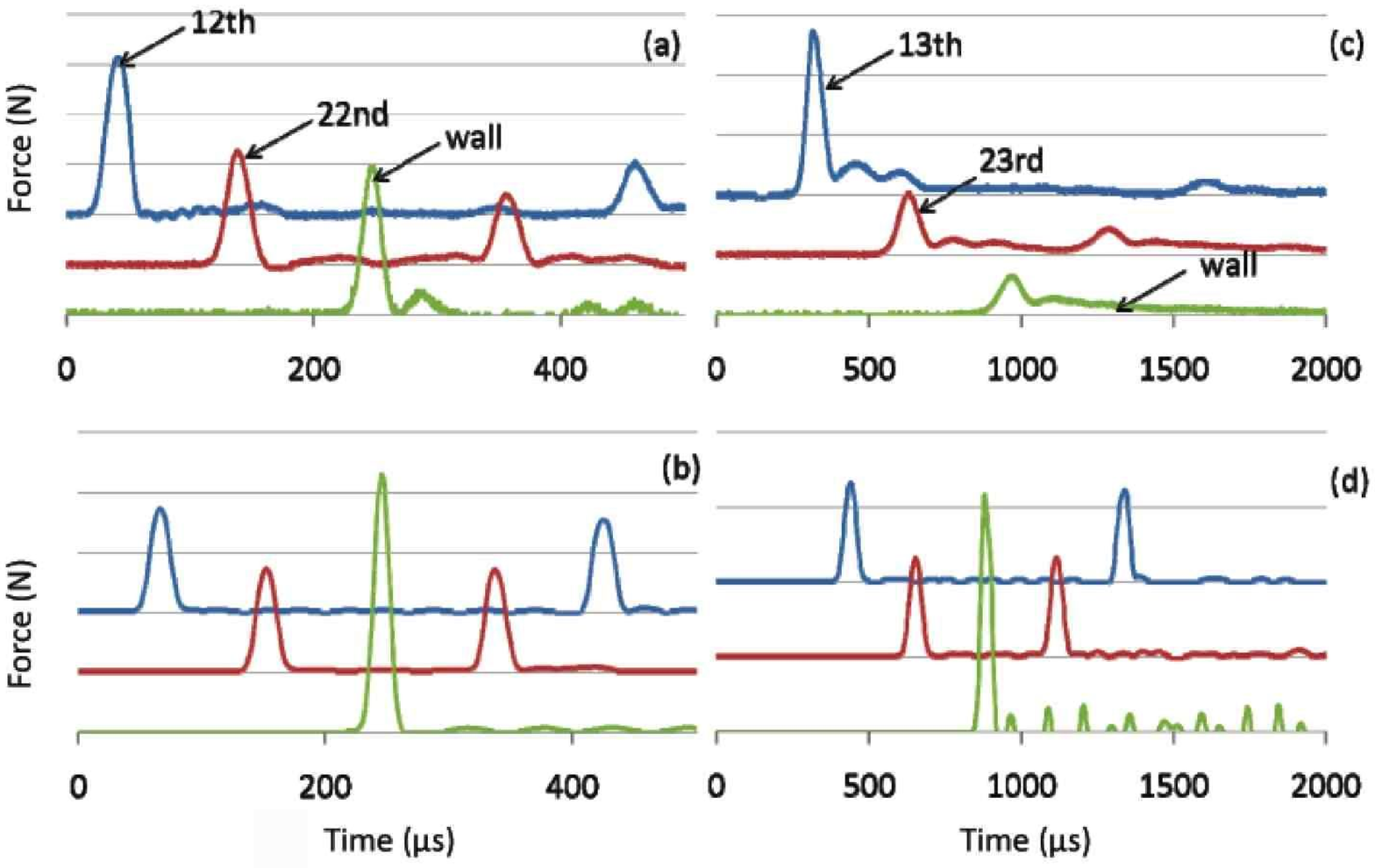}}
\caption{(Color online)  Force versus time response obtained from chains of dimers consisting of (a,b) 1 steel bead alternating with 1 bronze bead and (c,d) 1 PTFE bead alternating with 1 glass bead.  Panels (a,c) show experimental results, and (b,d) show the corresponding numerical data.  The numbered arrows point to the corresponding particles in the chains.  For the steel:bronze configuration, a steel striker impacted the chain with an initial velocity of 1.21 m/s.
The $y$-axis scale is 5N per divison in (a) and 20 N per division in (b).  
For the PTFE:glass configuration, a glass striker impacted the chain with an initial velocity of 1.17 m/s.
The $y$-axis scale is 1 N per division in (c) and 3 N per division in (d).  
}
\label{oneone2}
\end{figure}

We begin by discussing our results for 1:1 dimers.  In Fig.~\ref{oneone}, we show experimental and numerical results for $1:1$ dimers of steel:PTFE [panels (a,b)] and steel:rubber [panels (c,d)] particles.  The steel:PTFE chain consisted of 38 beads, and the steel:rubber chain had 19 beads; in each case, we used a steel bead as the striker.  The dynamics indicate that the initial excited impulse develops into a solitary wave within the first 10 particles of the chain.   Observe in the steel:PTFE numerics the presence of small-period oscillations after the large solitary pulse.  
These arise as residual ``radiation'' emitted by the tail of the wave as it 
attempts to progress through the highly nonlinear lattice. This is 
reminiscent of radiative phenomena in different classes of nonlinear lattices,
as discussed, for example, in Ref.~\cite{peykru}.

Interestingly, even the chains composed of alternating steel and Neoprene rubber beads (inherently nonlinear elastic components) support the formation of solitary-like pulses.  However, this configuration is highly dissipative and the solitary waves are consequently short-lived (although this feature is not incorporated in the theoretical model considered herein).  As shown in Fig.~\ref{oneone}(c,d), the initially very short pulse is quickly transformed into a much wider and slower pulse compared to the dimer composed of steel and PTFE particles.  Additionally, the presence of dissipation in the steel:rubber configuration tends to dampen the propagation of the pulses after the first 15 beads.  Incorporating dissipative effects is a natural direction for the refinement of the model to be pursued in future studies.  

To illustrate the robustness of solitary-wave formation, we also studied dimer chains composed of other types of materials, including steel:bronze (33 total particles), PTFE:glass (34 particles), and PTFE:nylon (33 particles) configurations.  Two important properties of the pulses observed in these systems are their propagation speeds and widths (measured by the full width at half maximum, or FWHM).  We compute the pulse speed using time-of-flight measurements for both experiments and computations.  To do this, we first measure the peak force of the pulse at each bead and determine the times at which these peaks occur.  We then estimate the velocity of the pulse at each bead by examining how long it takes for the pulse (as measured at the peak time) to travel from an earlier bead to a later one.  For the numerical calculations, we use the same bead separation distance as in corresponding experiments.  For example, for the 1:1 steel:PTFE dimer chain experiments, forces were measured at beads positions 13 and 24, so in the numerics we use time-of-flight measurements with an interval of 11 beads.  

We depict the results for steel:bronze in Fig.~\ref{oneone2}(a,b) and for PTFE:glass in Fig.~\ref{oneone2}(c,d).  The steel-bronze dimer chain included 38 total particles; its striker was a steel bead, which was dropped from about 9.5 cm.  We investigated this chain in order to examine a configuration with two different types of large-mass materials with relatively large elastic moduli.  It is evident from both experiments and numerical simulations that the system supports the formation and propagation of highly nonlinear solitary waves.  These waves have a pulse speed of 499.2 m/s experimentally and 544.0 m/s numerically.  Their FWHM was measured at 2.03 beads and 2.06 beads in experiments and numerics, respectively.  PTFE:glass dimer chains provide another configuration consisting of two different particles with similar masses, but in this case the masses are smaller than with steel:bronze and the elastic moduli and (especially) Poisson ratios of the two materials are rather different.  Although it is composed of lighter and softer particles, this dimer chain nevertheless supports the formation and propragation of solitary-like pulses.  However, many of their properties are quite different: the signal speed is significantly reduced (it is 151.7 m/s in experiments and 221.4 m/s in the numerics) and the pulse FWHM is 1.99 experimentally and 2.10 beads numerically.  In addition, secondary oscillations can be observed both in experiments and (to a lesser extent) in numerical simulations [see, in particular, the bottom (wall) signal in Fig.~\ref{oneone2}(d)].  Finally, the PTFE:nylon chain (with a PTFE striker dropped from a height of 2.5 cm) provides a configuration with one small mass and a second mass that is extremely small.  Unlike rubber, the nylon beads do not lead to severe dissipation.  However, this configuration appears to have greater difficulty in forming and supporting solitary waves, at least in the chain lengths studied (33 beads).  

As we have suggested, dimer chains have decidedly different properties from uniform chains.  We find numerically that the solitary pulse in uniform chains has an FWHM of about 2.06 cells (i.e., 2.06 beads), consistent with prior observations (see, e.g., Fig.~1.4 in \cite{nesterenko1}).  Because of the similar masses in the steel:bronze and PTFE:glass nylon chains, the pulse width in those 1:1 dimer chains is similar to that obtained for uniform chains.  We observe this feature both experimentally and numerically.  (As discussed later, the theoretical prediction for the mean FWHM in dimer chains depends only on the mass ratio of the two bead materials; the existence of the dimer is built into the long-wavelength asymptotics used to derive this result.)  This suggests that there is a ``critical" mass ratio $m_1/m_2$ necessary for the system to become ``sensitive" to the existence of the dimer.  This poses an interesting question concerning what is the critical point (mass ratio) for the dynamics to resemble that of a chain with periodic ``defects."  That is, how large (or small) should $m_2$ be to obtain bona fide dimer dynamics (increased pulse width, etc.)?
 
In the case of homogeneous chains, asymptotic analysis predicts a wave width of about 5 cells from tail to tail \cite{nesterenko1}.  The propagation speed of the pulses in steel is quite large; for steel strikers with an impact velocity of 1.37 m/s, we found numerically that the pulse propagates at almost 700 m/s, and other authors have found large experimental propagation speeds with other initial conditions \cite{coste99}.  In contrast, as discussed in detail below, pulses in dimers have a shorter width in terms of number of cells.  
With a steel striker with a velocity of 1.37 m/s on impact, we find a propagation speed of 168.1 m/s numerically and 127.8 m/s experimentally.  We discuss the width and pulse propagation speed of solitary waves in more detail below and include an asymptotic analysis of the width of 1:1 dimers as compared to the previously-studied limiting cases of monomer chains and 1:1 dimers with beads of mass $m_1$ and $m_2 \ll m_1$ \cite{nesterenko1}.  Preliminary experiments and numerical simulations on 1:1 dimers, considering only materials with elastic moduli of the same order of magnitude (and mass ratios $m_1/m_2$ of 2, 4, 16, 24, and 64), were described in Ref.~\cite{nesterenko1}.

\subsection{$N_1:1$ and $1:N_2$ Dimers}

\begin{figure}[tbp]
\centering \includegraphics[width=15.0cm]{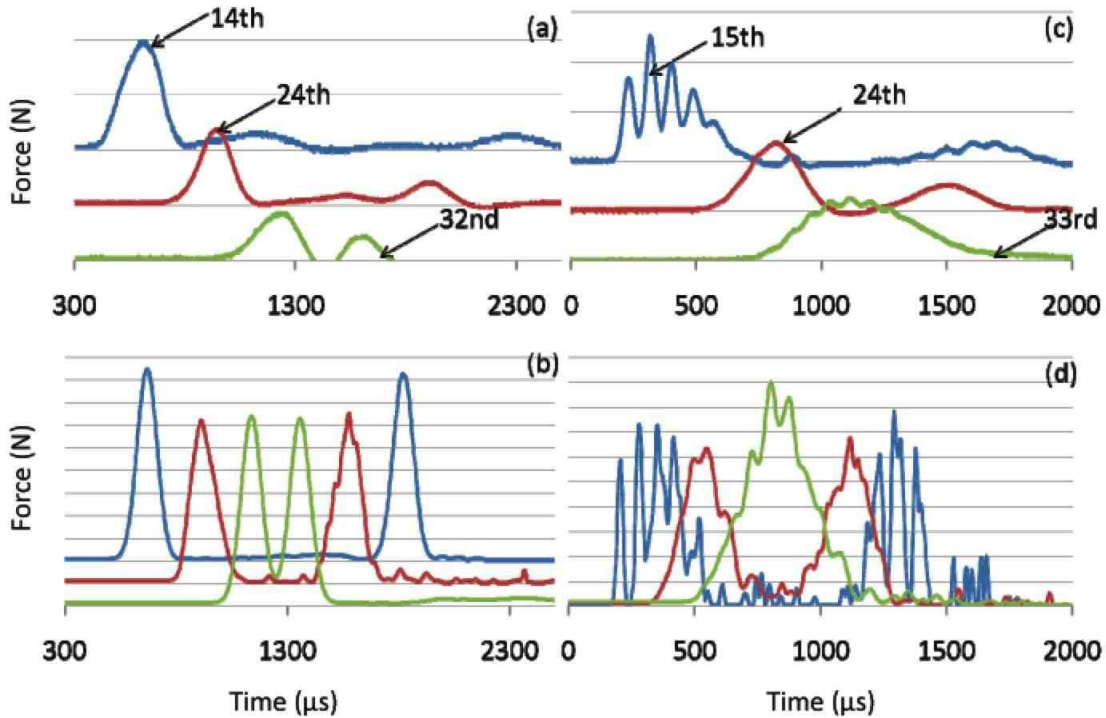}
\caption{(Color online) Force versus time response obtained from chains of dimers consisting of (a,b) 2 and (c,d) 5 stainless steel beads alternating with 1 PTFE bead.  The total number of beads is 38 in each case.  Panels (a,c) show experimental results, and (b,d) show the corresponding numerical data.  For both experimental configurations, the striker had an impact velocity of 1.37 m/s with the chain.
The $y$-axis scale is 2 N per division for (a,c) and 1 N per division for (b,d).  The numbered arrows point to the corresponding particles in the chain.  For both configurations, the second curve (showing the results for particle 24) represents a PTFE bead and the other curves represent steel beads.
}
\label{none}
\end{figure}

\begin{figure}[tbp]
\centering \includegraphics[width=15.0cm]{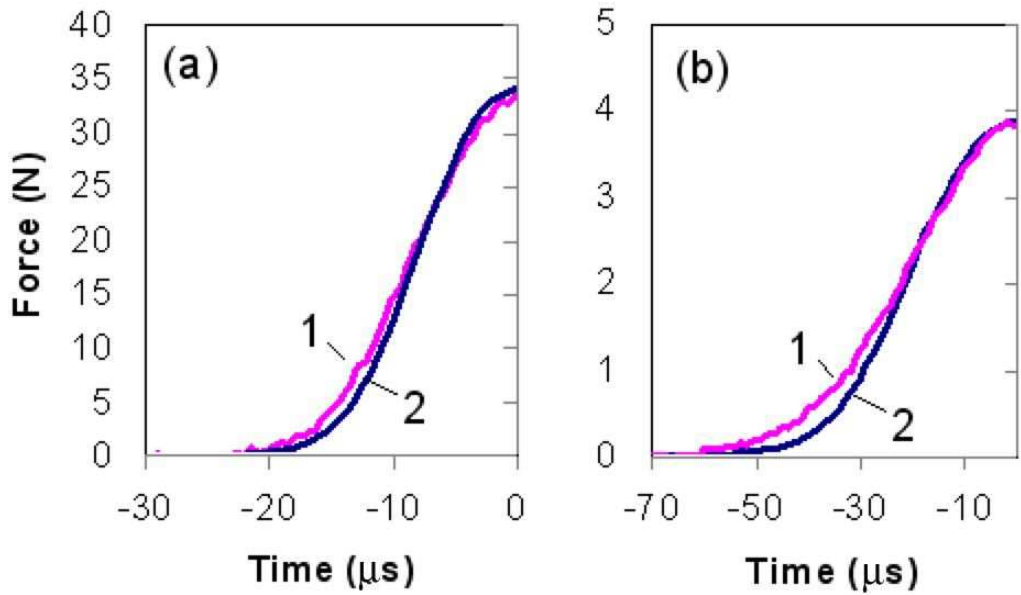}
\caption{(Color online) Rising shape comparison of two different solitary waves recorded in the middle of the chain for different dimer configurations.  Each panel displays the evolution of the force (in N) with time (in microseconds). Curve 1 corresponds to the experimental points, and curve 2 represents the numerical values. The depicted examples are (a) steel-bronze dimer curves detected in particle 12 and (b) PTFE-glass dimer curves detected in particle 12.  Note that in order to compare the shapes, the experimental curves have been multiplied by an arbitrary factor to obtain the same force amplitude as that found in the numerical data (where no dissipation is present). 
}
\label{direct}
\end{figure}

\begin{figure}[tbp]
\centering \includegraphics[width=15.0cm]{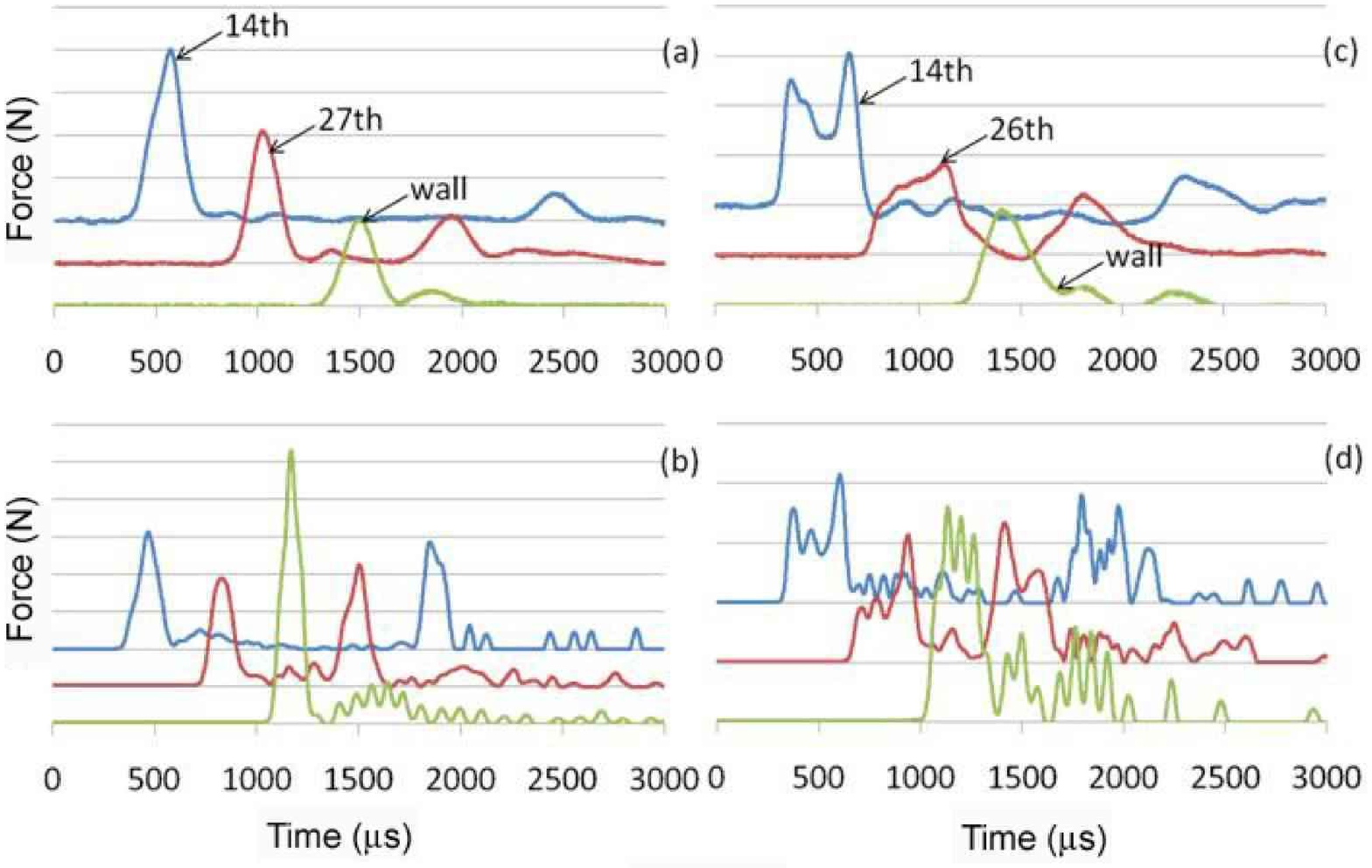}
\caption{(Color online) Force versus time response obtained from chains of dimers consisting of 1 steel bead alternating with (a,b) 2 and (c,d) 5 PTFE beads.  In each case, the total chain length was 38 particles.  Panels (a,c) show experimental results, and (b,d) show the corresponding numerical data.  For both configurations, the striker (a steel bead) had an impact velocity of 1.37 m/s with the chain.
The $y$-axis scale is 1 N per division in (a,c) and 2 N per division in (b,d).  The numbered arrows point to the corresponding particles in the chain.  In (a,c), both particle 14 and particle 27 are made of PTFE.  In (b,d), particle 14 is PTFE and particle 26 is steel.  
}
\label{onen}
\end{figure}

We consider configurations with different periodicities by varying the number of steel and PTFE particles in steel: PTFE dimers.  We obtain robust pulses for $N_1:1$ dimers with $N_1 >1$ (with $N_1$ as high as 7 in experiments and as high as 22 in numerics), though the transient dynamics and spatial widths of the developed solitary-like waves are different. This is clearly illustrated
in Fig.~\ref{none} through experimental and numerical results for $2:1$ and $5:1$ steel:PTFE dimers.
The difference in transient dynamics can be intuitively understood in terms of the increasing
``frustration'' of the system as $N_1$ increases (due to the presence of the 1 bead of 
different type in each cell of an otherwise uniform chain). On the other hand, the spatial widths differ
through a mechanism similar to the one that we will explain analytically (see the discussion below) for the $1:1$ configurations.
Each bead in a given cell responds differently but consistently throughout the chain. For 
example, in the 2:1 chain, the second steel bead in a cell possesses a cleaner and more 
solitary-like shape than the first or third bead.  This is especially evident in the numerical 
simulations.  

For the 5:1 dimers, there is a critical ``frustration" between the pulse size one would obtain 
in a uniform chain of beads (for which the predicted width is $\approx 5$ particles long \cite{nesterenko1}) versus that one would obtain in a 5:1 chain (about 30, corresponding to the presence of 5 unit cells with 6 particles each.  The pulses are of smaller (spatial) width in terms of the number of cells than they are in the 1:1 dimers.  More generally, we observed both numerically and experimentally that the width (in terms of the number of cells) of the solitary waves decreases gradually with increasing numbers $N_1$ of steel beads in a single dimer (cell) from the 5-cell width expected for Hertzian interactions.  It takes roughly six cells to achieve a stable pulse for the 2:1 configuration, whereas it takes 12 cells for the 5:1 configuration.  (It also takes roughly 6 cells for the pulse in the 3:1 chain to stabilize and roughly 11 cells for that in the 4:1 chain to stabilize.)  The increased propagation distance needed to stabilize the pulse is a result of the frustration discussed above.  This frustration reflects the increased difficulties in considering an $N_1:1$ cell as a ``quasiparticle" as $N_1$ increases.

We observe in our experiments that the width of the solitary waves (in terms of the total number of cells) decreases, although it increases in terms of number of ``participating'' beads.  For the numerics, the FWHM seems to be steady at first, but it is smaller for $N_1 = 4$ as the ``frustration" arises.  We observe that the pulse speed increases with increasing $N_1$ in both experiments and numerics.   This is physically justified because the increase in the number of steel particles makes the system generally ``stiffer."  For each configuration, the scaling between the pulse's maximum force and its speed of propagation is $V_s \sim F_M^{1/6}$, whose validity we verify analytically for 1:1 dimers using long-wavelength asymptotics (see below).  We compare the shapes of the experimental and numerical pulses at equilibrium for example 1:1 chains in Fig.~\ref{direct}.

We performed analogous experiments and numerics for relatively short chains consisting of 38 particles assembled in a $1:N_2$ setup (i.e., 1 stainless steel and $N_2 > 1$ PTFE particles).  We show the results for $N_2 = 2$ in Fig.~\ref{onen}(a,b) and those for $N_2 = 5$ in Fig.~\ref{onen}(c,d).  For $N_2 = 2$, we again observed solitary waves, though some additional smaller amplitude 
structures were also present.  For $N_2 = 5$, we did not obtain robust, individual solitary pulses in either experiments or numerical simulations of short chains but instead observed a complex ``rattling" between the particles, which resulted in the formation of trains of pulses of different amplitudes.  Some structure was more apparent in the longer $1:N_2$ chains we simulated numerically, but we still did not obtain robust solitary waves (see the discussion below).

To confirm that robust solitary waves form even for dimers with larger numbers of steel particles in a cell (i.e., large $N_1$ in $N_1:1$ steel:PTFE dimer chains) despite the ``frustration" phenomena observed above, we conducted additional numerical investigations with a large number of particles.  
We used horizontal chains of beads (setting $g = 0$) to avoid the ``linearizing" effects  of gravity \cite{nesterenko2,hong01b} 
due to the significant gravitational precompression arising from the large number of particles.  

\begin{figure}[tbp]
\centering{(a)\includegraphics[width=7.5cm]{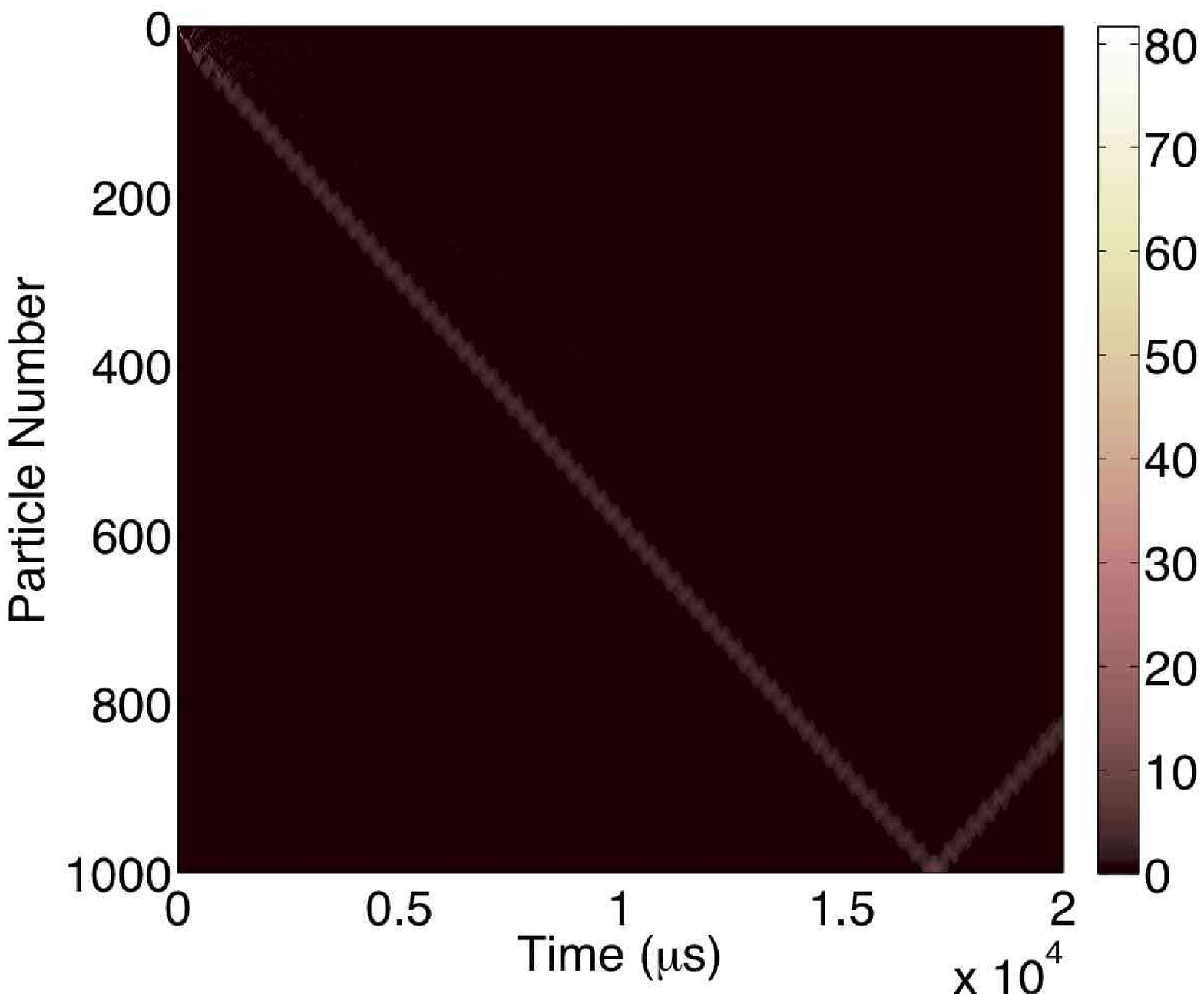}
(b)\includegraphics[width=7.5cm]{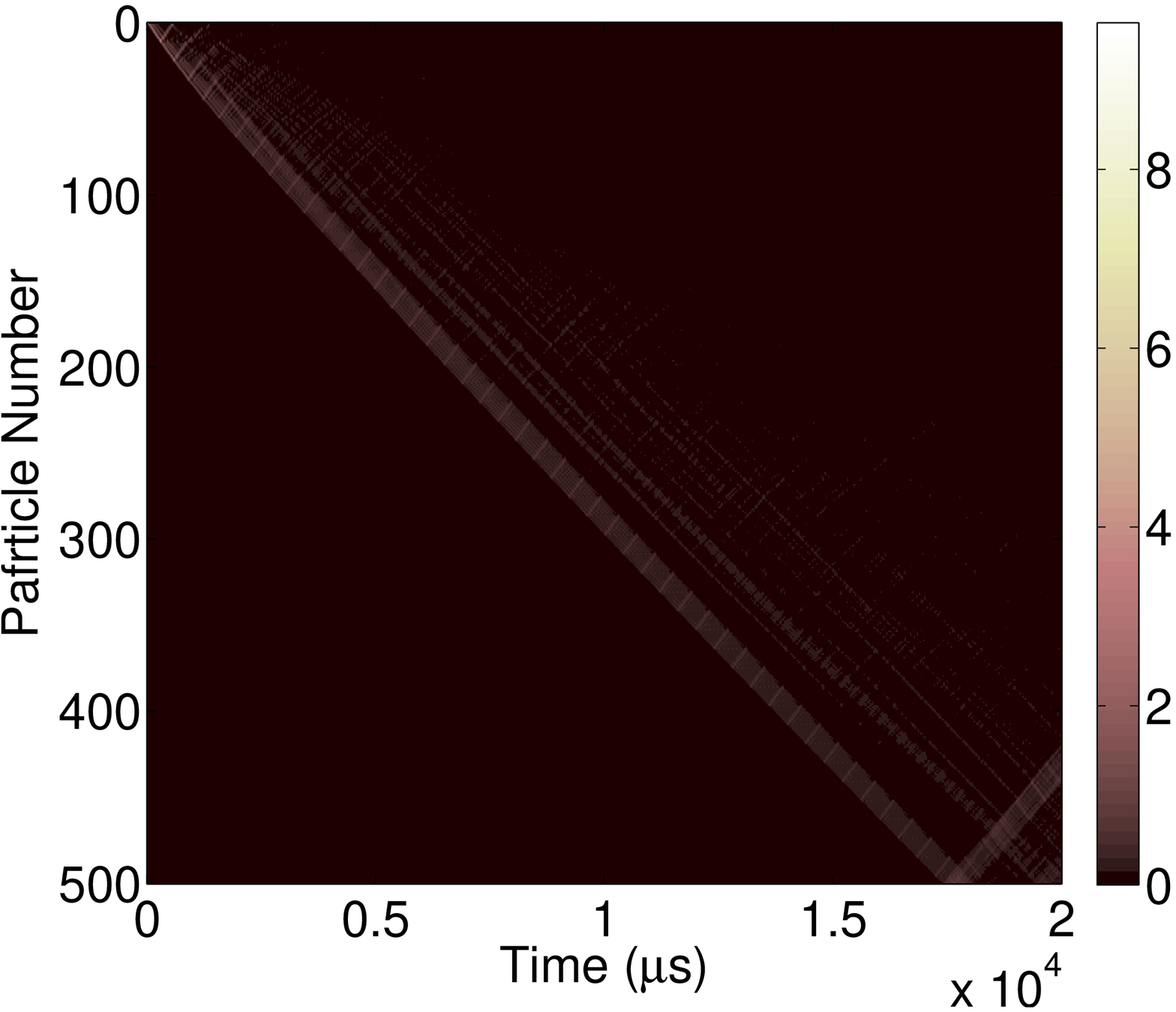}}
\caption{(Color online) Space-time diagrams for long chains of (a) $15:1$ and (b) $1:10$ steel:PTFE dimers.  For both configurations, the striker had a velocity of 1.37 m/s on impact.  The striker was made of steel for the 15:1 dimer and made of PTFE for the 1:10 dimer.  
}
\label{longchains}
\end{figure}

In panel (a) of Fig.~\ref{longchains}, we show a space-time plot for a 15:1 steel:PTFE chain of 1000 particles.  As the figure demonstrates, one gets well-formed localized pulses even for larger $N_1$ when the chain is sufficiently long.  However, the shape of the pulse never stabilizes fully, as there are fluctuations in the peak as well as more extensive fluctuations next to the peak than what one sees near the peaks for smaller $N_1$ (which do stabilize).  Panel (b) of Fig.~\ref{longchains} shows a space-time plot for a 1:10 steel:PTFE chain of 500 particles.  While a larger peak is clearly observable, one does not obtain a pulse that is significantly larger than its surrounding peaks the way one does for dimers with several consecutive large-mass particles.  We thus conclude that dimers with several consecutive large-mass beads behave rather differently in certain respects than dimers with several consecutive small-mass particles.  This asymmetry would be interesting to study systematically in the future, as it might provide insights into considering the lattice chain discretely versus as a continuum.  For example, we expect different excitation modes to be excited in lighter versus heavier particles.

\subsection{Theoretical Analysis} 

In our theoretical considerations, we investigate the prototypical dimer chain composed of 
$1:1$ dimer cells containing beads with different masses (denoted by $m_1$ and $m_2$).   We derive for the first time the long-wavelength equation for an arbitrary mass ratio and a general interaction exponent $k$.  (We included a brief synopsis of this derivation in Ref.~\cite{dimer}.)  The relevant rescaled dynamical equations can be expressed as 
\cite{nesterenko1}
\begin{align}
	m_1 \ddot{u}_j &= (w_j-u_j)^k-(u_j-w_{j-1})^k\,, \label{eq1} \\
	m_2 \ddot{w}_j &= (u_{j+1}-w_j)^k - (w_j-u_j)^k\,, \label{eq2}
\end{align}
where $u_j$ ($w_j$) denotes the displacement of the 
$j$th bead of mass $m_1$ ($m_2$). We present our
analysis for the nonlinear wave in the case of general power-law 
interactions to illustrate the generality of the approach.  We subsequently use Hertzian contacts (i.e., the special case of $k = 3/2$) in order to compare our 
theoretical analysis with our experimental results.  
The $k = 1$ case corresponds to linear interactions between particles.


The distance from $u_j$ to $w_j$ (from $u_j$ to $u_{j+1}$) is denoted by $D$ 
($2 D$).  This gives a small parameter, allowing to us develop a
long-wavelength approximation (LWA) by 
Taylor-expanding Eqs.~(\ref{eq1})-(\ref{eq2}). In particular, 
we express $u_{j+1}$ ($w_{j-1}$) as a function of $u_j$ ($w_j$). 
The resulting PDEs (one for each mass ``species'') need to be 
``homogenized.'' To accomplish this, we follow \cite{pnevmatikos}, 
postulating a consistency condition between the two fields:
\begin{align}
	w & =\lambda \left(u + b_1 D u_x + b_2 D^2 u_{xx} + b_3 D^3 u_{xxx} + b_4 D^4 u_{4x}
+ \dots \right)\,. \label{eq3}
\end{align}
We then self-consistently determine the coefficients $\lambda$ and $b_i$ by demanding that Eqs.~(\ref{eq1}) and (\ref{eq2}) be identical {\it at each order}.
Note that the subscripts in the LWA denote derivatives.  
For some technical issues regarding the consistency condition, 
we defer the interested reader to the discussion in \cite{pnevmatikos}
[see, in particular, Eqs. (3.2)-(3.7) and the accompanying discussion]. 
The parameter $\lambda$ can take the values $1$ (for acoustic excitations) or 
$-m_1/m_2$ (for optical ones).  Given that our 
initial excitation {\it generically} 
produces in-phase (acoustic) waveforms, we restrict our 
considerations to $\lambda=1$ in the following exposition.
Observe that for optical (out-of-phase) excitations to be produced, a fundamentally
different and more complex experimental setup is necessary (in comparison to
the experiments discussed herein) to be able to drive the
system at a frequency within the gap of its linear spectrum.  

A direct comparison of the resulting PDEs for the two mass types 
at orders $D^k$--$D^{k+3}$ result in the following consistency constraints:
\begin{align}
	b_1 &= 1\,, 
\label{beq1}
\\
	b_2 &= \frac{m_1}{m_1+m_2}\,,
\label{beq2}
\\
	b_3 &= \frac{2 m_1-m_2}{3 (m_1+m_2)}\,,
\label{beq3}
\\
	b_4 &= \frac{m_1 (m_1^2-m_1 m_2 + m_2^2)}{3 (m_1+m_2)^3}\,.
\label{beq4}
\end{align}
We note in passing that these results accurately capture the uniform-chain limit of $m_1=m_2$. 
In this case, $w_j$ becomes effectively $u_{j+1}$ and hence the 
consistency condition of (\ref{eq3}) becomes the Taylor 
expansion of $u_j$.  The quantity $D$, which equals the sum of the radii of consecutive beads in the case of dimer chains, is the radius of this expansion.

The resulting PDE bearing the leading-order discreteness corrections (i.e., incorporating
terms that are $D^2$ below the leading-order term) is of the form
\begin{align}
	u_{\tau\tau} &= u_x^{k-1} u_{xx}+ G u_x^{k-3} u_{xx}^3 + H u_x^{k-2} u_{xx} 
u_{xxx}  + I u_x^{k-1} u_{4x}\,,
\label{eq4}
\end{align}
where $\tau=t \sqrt{2 k D^{k+1}/(m_1+m_2)}$ is a rescaled time.  The constants arising in Eq.~(\ref{eq4}) are given by 
\begin{align}
	G &= D^2 \frac{(2-3 k + k^2)m_1^2}{6 (m_1+m_2)^2}\,,
\label{beq5}
\\
	H &= D^2 \frac{2 (k-1) (2 m_1^2+m_1 m_2-m_2^2)}{6 (m_1+m_2)^2}\,,
\label{beq6}
\\
	I &= D^2 \frac{2 (m_1^2-m_1 m_2+m_2^2)}{6 (m_1+m_2)^2}\,.
\label{beq7}
\end{align} 

Given the nature of the excitation propagating through the chain of
beads, we seek traveling wave solutions $u\equiv u(\xi)$, where 
$\xi=x-V_s t$ is the standard traveling wave variable with (renormalized) speed $V_s = \frac{dx}{d\tau}$. This ansatz results in an ordinary differential equation (ODE) for $u_{\xi}=v$ [this
change of variable reduces the equation to third order]. In all the integrations that follow, we consider solutions with homogeneous Dirichlet boundary conditions far from the wave.

We now need to use some tools from the theory of ODEs to further
reduce the order of the equation. In particular, we
change variables using $v=z^p$, where the power $p$ is 
chosen so that terms proportional to $z^{p-3} z_{\xi}^3$ disappear
for $z = z(\xi)$:
\begin{equation}
	p_{1,2}=\frac{(H+3I) \pm \sqrt{(H+3 I)^2-8 I (G + H + I)}}{2 (G + H + I)}\,. \label{beq8}
\end{equation}
Observe that there are two possible values of $p$ that lead to this
reduction, but this does not change the essence of the results in
what follows. We obtain an ODE of the form
\begin{align}
	V_s^2 p z^{p-1} z_{\xi} &= p z^{kp-1} z_{\xi} + p I z^{k p -1} z_{\xi \xi \xi}
+ \left[3 p (p-1) I + p^2 H \right] z^{kp-2} z_{\xi} z_{\xi \xi}\,.
\label{beq9}
\end{align}  
In Eq.~(\ref{beq9}), we use an integrating factor 
$z^a$, with 
\begin{equation}
	a=1-k p + 3 (p-1) + p H/I\,, \label{beq9a}
\end{equation}
to convert it to a tractable second order ODE,
\begin{align}
	z_{\xi\xi}=\mu z^{\eta} - \sigma z \,, \label{eq5}
\end{align}
where $\mu=V_s^2/[I (p+a)]$, $\eta=1+p-k p$, and $\sigma=1/[I (kp + a)]$.  
One can find exact periodic solutions of Eq.~(\ref{eq5}) that take the form
\begin{align}
	u_{\xi} \equiv v \equiv z^p= B \cos^{\frac{2}{k-1}}(\beta \xi) \,, \label{eq6}
\end{align}
where $B=\left(\mu/[\beta^2 s (s-1)]\right)^{1/(k-1)}$, $\beta=\sqrt{\sigma} (1-\eta)/2$, and $s = 2/(1 - \eta)$.  

The coexistence of such an exact trigonometric solution with  nonlinear 
dispersion in the LWA of our model suggests the potential existence of 
essentially
finite-width solutions that consist of a single arch of the 
profile of Eq.~(\ref{eq6}) \cite{nesterenko1}.  Two key experimentally testable properties of such solutions are as follows:
\begin{itemize}
	\item{The amplitude-velocity scaling $B \sim V_s^{2/{(k-1)}}$.
Observe that this scaling does not change from the single-species 
case \cite{nesterenko1,nesterenko2}.  Accordingly, we conclude that this is a geometrical result that arises from the power of the nonlinearity (i.e., from the geometry of the contacts between consecutive particles).} 
	\item{The solution width that is equal to
$\pi/\beta$ and which, consequently, depends directly on the mass ratio.}  
\end{itemize}
Among these two properties, the width is the one that
naturally showcases the relevance and novelty of the general 
dimer theory developed herein in comparison
with the monomer theory that can be found in \cite{nesterenko1}.
The advantage of the present theory is that it allows us to obtain
an explicit, closed-form expression for the width that is valid for all $k$.
Although the formula is too lengthy to write explicitly in the general case, we present here the
expression for $\beta$ in the experimentally relevant case of $k=3/2$: 
$\beta=(30-8 \omega+8 \omega^2)^{-1} (\sqrt{3} (8-5 \omega+5 \omega^2-\sqrt{4-4\omega+13\omega^2-18 \omega^3+9 \omega^4})
(((1+\omega)^2(15-4\omega+4\omega^2))/(D^2 (34-42 \omega+59 \omega^2-34
\omega^3+17 \omega^4+(-8+5 \omega-5 \omega^2)\sqrt{4-4\omega+13\omega^2-
18 \omega^3+9 \omega^4})))^{1/2})$, where $\omega=m_2/m_1$.  
One of the appealing features of this result is that it naturally 
generalizes all previously-known limiting cases -- namely, 
the monomer case ($m_1=m_2$), for which $\beta=\sqrt{10}/(5 D)$ [resulting in pulses extending to 
$5\pi /\sqrt{10} \approx 5$ sites], and the case with one mass 
that is much larger than the other ($m_1 \gg m_2$) [resulting in 
pulses of $\sqrt{10} \pi \approx 10$ sites, consisting of 5 cells 
with 2 sites each] \cite{nesterenko1}.

\begin{figure}[tbp]
\centering \includegraphics[width=150mm]{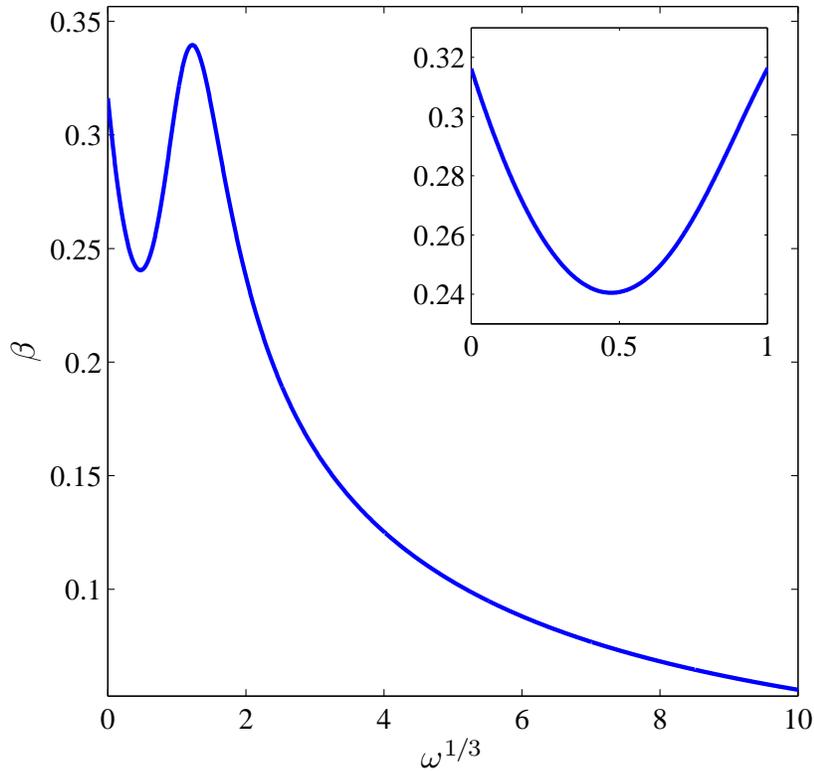}
\caption{The parameter $\beta$ as a function of the dimensionless mass-ratio parameter $\omega^{1/3}$ for a system with $k = 3/2$, $\rho_{2}/\rho_{1}=1$, $R_{1}=1$, and $R_{2} \in [0, 10]$.  Note the approximate local symmetry in the inset.}
\label{beta_fig1}
\end{figure}

Before presenting a comparison between our theoretical and experimental results, we wish to briefly highlight that there is an inherent asymmetry in the quantity $\beta$ in Eq.~(\ref{eq6}).  This quantity is scaled to the spatial wavelength of $D = R_1 + R_2$ and depends intrinsically on the ratio of the radii and masses of consecutive particles through the parameter combination $\rho_2/\rho_1 \times R_2^3/R_1^3$ (where $\rho_j$ is the density of particle $j$).  Consequently, one should not expect the resulting expression for $\beta$ to be symmetric about $\omega = 1$, so that it matters whether $R_1$ or $R_2$ is larger.  We show this asymmetry in Fig.~\ref{beta_fig1} for the case of Hertzian interactions.  Observe in the inset, however, an approximate local symmetry about $\omega^{1/3} \approx 0.5$ and the consequent consistency in the derived wavelengths of $10D \approx 10R_1$ at $\omega = 0$ (where $\beta = 1/\sqrt{10}$) and $5D = 5(R_1 + R_2) = 10R_1$ about $\omega = 1$ (where $\beta = 1/\sqrt{10}$ again).  This also brings up the question of when it is appropriate to employ continuum versus discrete treatments of the granular chains.


\subsection{Comparison between experiments, numerical simulations, and theory}

\begin{figure}[tbp]
\centering
\includegraphics[width=15.0cm]{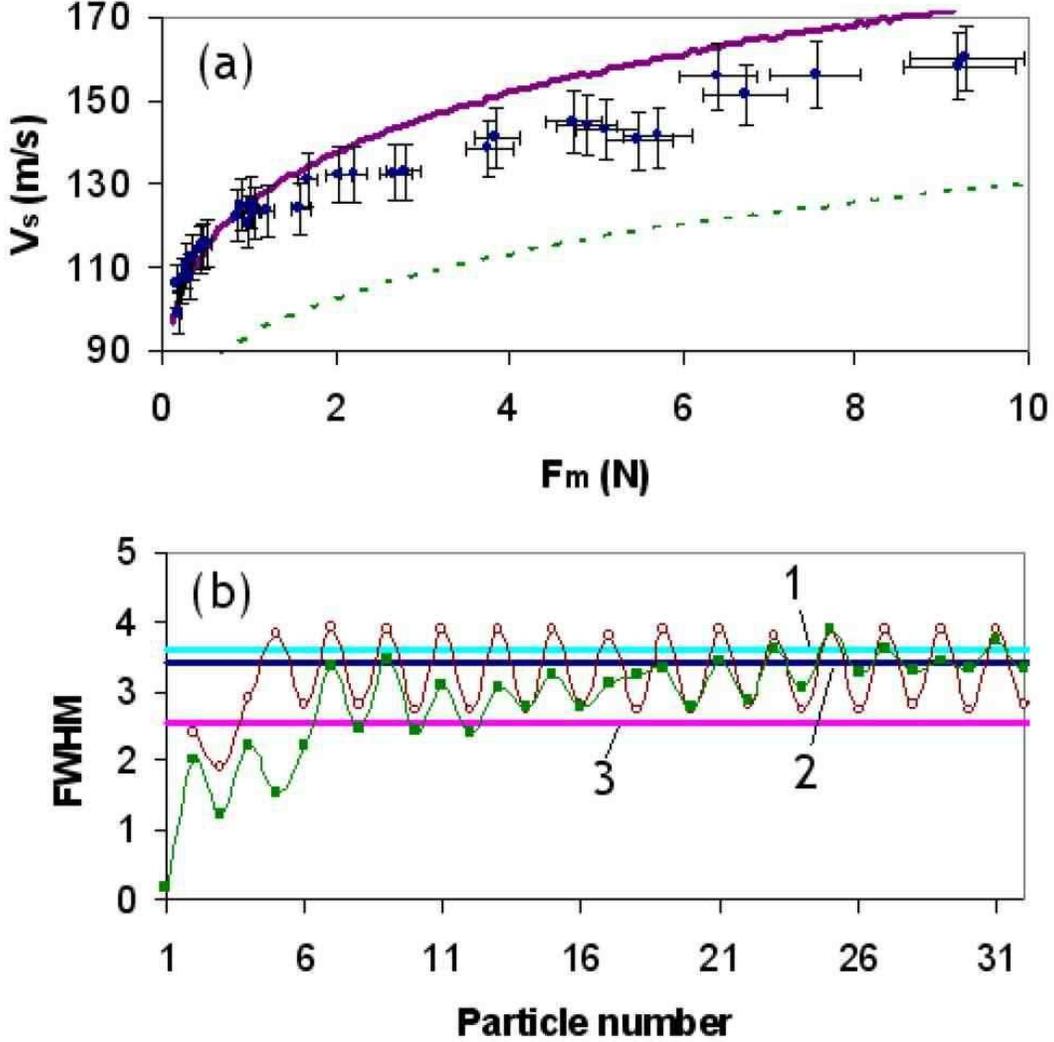}
\caption{(Color online) Comparison of experiments, numerical simulations, and theory.  (a) Scaling of the maximum dynamic force $F_{m}$ versus pulse propagation speed $V_s$ in a 1:1 stainless steel:PTFE dimer chain.  The numerical simulations are shown by the solid curve and the experimental results are shown by points (with error bars).  The dashed curve shows the numerical results using the elastic modulus $E = 0.6$ GPa for PTFE, the nominal static value reported in most of the literature \cite{dar05,dupont}.  (b) Evolution of solitary wave width (the FWHM) as a function of bead number.  The experimental values are shown by solid (green) squares and the numerical values are shown by open (red) circles.  (In both cases, we include curves between the points as visual guides.)  The theoretical value for the FWHM with $m_1 \gg m_2$ is given by line (1), that for the 1:1 steel:PTFE chain is given by line (2), and that for a homogeneous chain is given by line (3).
}
\label{compare}
\end{figure}

First, we discuss a relation for the scaling of the maximum dynamic force $F_{m}$ of the solitary waves versus their propagation speed $V_s$.  From the theory, we find that $F_{m} \sim B^k \sim V_s^{2k/(k-1)}$, which for $k = 3/2$ yields $V_s \sim F_{m}^{1/6}$, the same relation as for monomer chains \cite{nesterenko1}.  We tested this result both experimentally using 1:1 steel:PTFE dimer chains and numerically for both 1:1 steel:PTFE and 1:1 steel:rubber chains.  In all cases, the chains have 38 total beads.  In the experiments, we dropped a striker (a stainless steel bead) from heights between 0 m and 1.2 m and measured the peak force in beads 15 and 34.  We averaged their amplitudes [$F_{m} = (F_{15}+F_{34})/2$] and obtained the corresponding wave speed ($V_s$) using time-of-flight measurements.  We computed $F_m$ and $V_s$ in exactly the same manner in our numerical simulations and obtained good agreement, as indicated in Fig.~\ref{compare}(a).  We show error bars for the experimental results.  The errors in the force amplitude stem from oscilloscope noise and sensor orientation (their calibration varies a bit because they are aligned with the horizontal cut); those in the propagation speed arise from oscilloscope noise and the difficulty in determining exact time-of-flight distance between the peaks.  In conducting our numerical simulations, we performed a least-squares fit to the obtained force and velocity values.  To compare the numerical results with the theory, we repeated the numerical simulations in the absence of gravity, obtaining $V_s \sim F_{m}^{0.1666}$, which is in excellent agreement with the theory (even though we used a small number of particles).  We remark that the inclusion of gravity, which decreases the exponent $\gamma$ in $V_s \sim F_m^\gamma$, has a larger effect on the force-velocity scaling in dimer configurations that include very light particles (such as steel:rubber dimers).  We verified numerically that the same force-velocity scaling also holds for $N_1:1$ steel:PTFE dimer chains with larger $N_1$ and for $1:1:1$ trimer chains. 

We saw earlier that the FWHM of solitary waves in a dimer chain is larger in terms of number of particles (though smaller in terms of number of cells) than that in a monomer chain.  The physical mechanism behind this observation can be explained using the ``homogenized'' theory (that is, the long-wavelength asymptotics) derived above.  We demonstrate this by examining the evolution of the FWHM as the wave propagates down the chain of beads.  As shown in Fig.~\ref{compare}(b), the numerical and experimental results are in very good agreement with each other and with the long-wavelength approximation that captures the average of the FWHM oscillations.  The maximum dynamic forces ($F_m$) in the numerics and experiments alternate from one bead to another because consecutive beads are composed of different materials.  From Fig.~\ref{compare}(b), it is clear that the average FWHM of the dimer differs both from that of a homogeneous chain ($m_1 = m_2$) and from that of the $m_1 \gg m_2$ limiting case discussed in \cite{coste97}. 
The average value of this width is best captured (for the proper ratio of masses) by the homogenized theory derived above, clearly evincing the relevance of our theoretical  approach.  To provide a quantitative measure of the theory's accuracy, we compute the relative error of the theoretical
prediction versus the computational-average FWHM and compare it to those for the two limiting cases; we find errors of $22.8\%$ for $m_1=m_2$, $9.2 \%$ for $m_1 \gg m_2$, and $2.5 \%$ for the correct mass ratio.  The numerical simulations achieve a nearly-steady FWHM value by about the 4th bead, and the experiments achieve this by about the 6th bead, so the theoretical value for the width of the solitary wave becomes valid very early in the wave propagation dynamics. 
 
We now comment briefly on the relative importance of gravitational effects for chains of beads with small and large numbers of particles.  First, because of the chain lengths and the pulse amplitudes considered in our experimental configurations, the pulse propagation can be safely assumed to be within the highly nonlinear regime.  We consequently neglected the effects of gravity in the theoretical analysis, in accordance with the discussion of Ref.~\cite{hong01b}.  However, for vertical chains of beads that contain a large number of particles (several hundred or more) or are excited by smaller-amplitude pulses, the presence of the nonuniform gravitational precompression should be taken into account.  As discussed in Ref.~\cite{hong01b}, this leads to a 1/3 power-law scaling in the wave width as a function of particle number as well as a power-law drift in the pulse speed.


\section{Chains of Trimers}

\begin{figure}[h]
\centerline{\includegraphics[height=7.5cm]{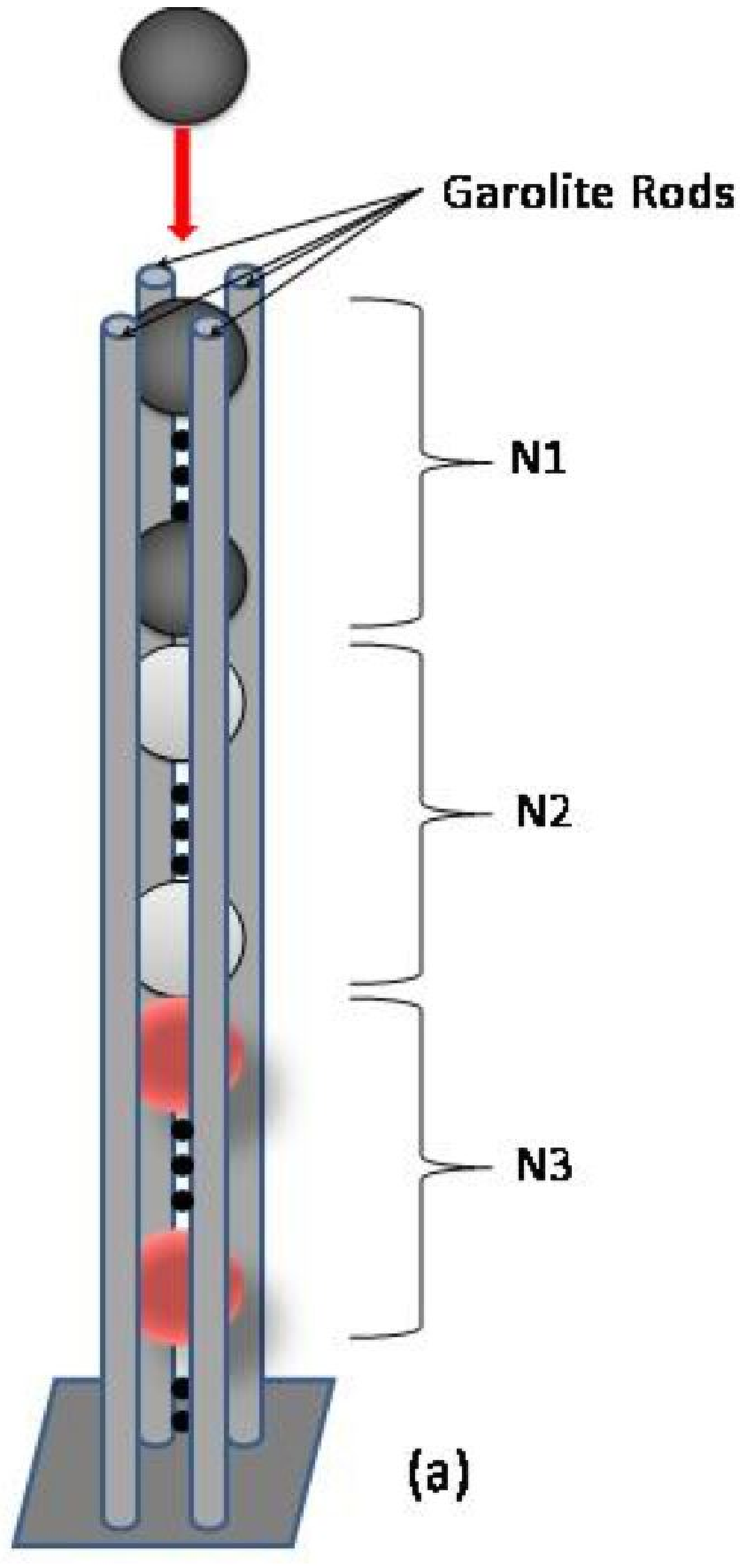}
\includegraphics[height=7.5cm]{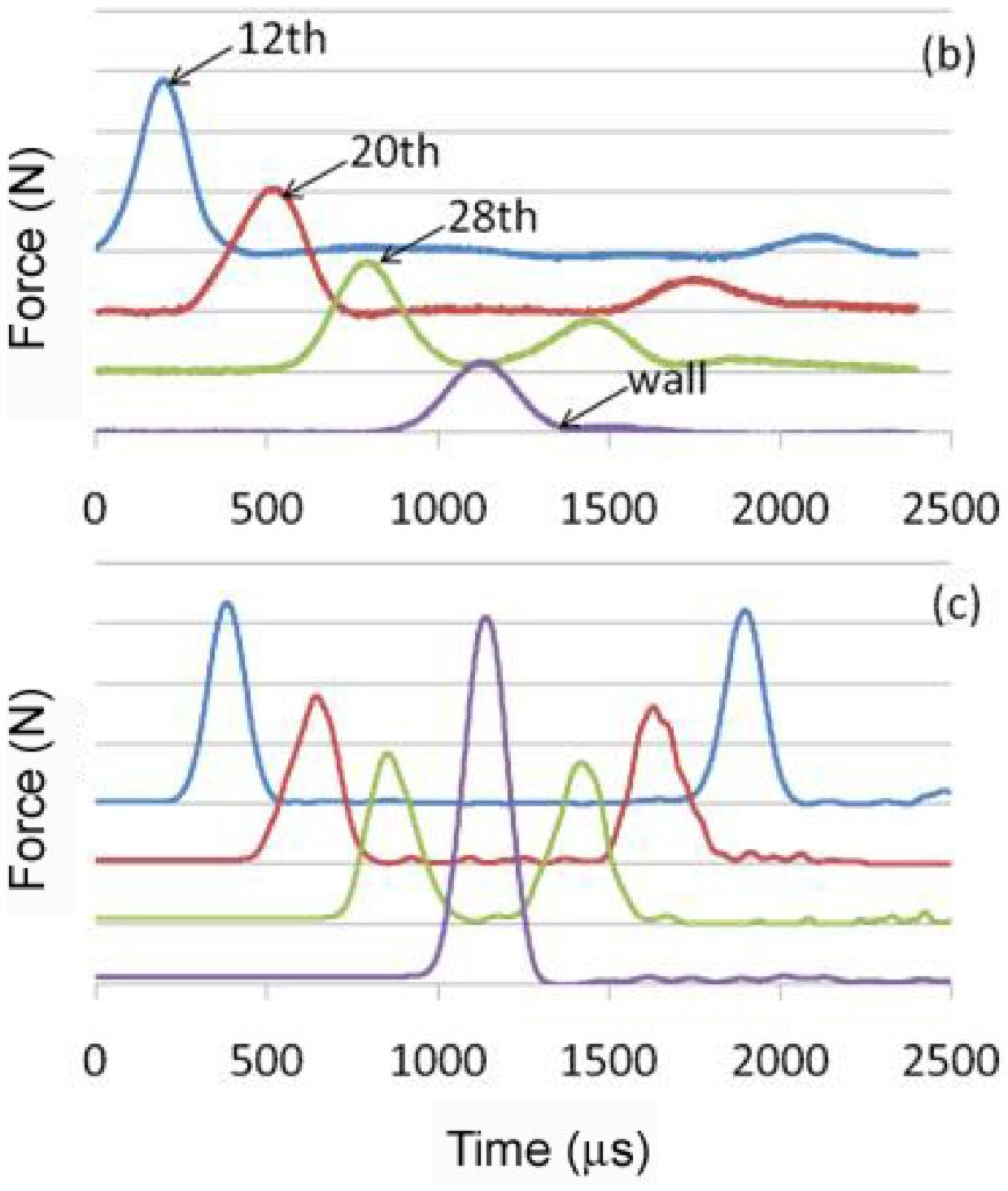}}
\caption{(Color online) (a) Experimental setup for trimer chain consisting of a periodic array of $N_1$ consecutive beads of one material, $N_2$ consecutive beads of a second material, and $N_3$ consecutive beads of a third material.  (b,c) Force versus time response obtained from chains of trimers consisting of 1 steel particle, 1 bronze particle, and 1 PTFE particle.  Panel (b) shows experimental results and (c) shows numerical ones.  A steel striker was dropped with a velocity of 0.54 m/s on impact,
and the $y$-axis scale is 1 N per division in both plots.  The numbered arrows point to the corresponding particles in the chain.  The 12th particle is PTFE, the 20th particle is bronze, and the 28th particle is steel.
}
\label{oneoneone}
\end{figure}

To examine more intricate heterogeneous chains, we also performed experiments and numerical simulations for various trimer configurations.  In Fig.~\ref{oneoneone}(a), we depict the basic experimental setup of a trimer chain in which each cell consists of $N_1$ consecutive beads of one type, $N_2$ consecutive beads of a second type, and $N_3$ consecutive beads of a third type.  We focused on $N_1 = N_2 = N_3 = 1$ and considered various material components with different masses, elastic moduli, and Poisson ratios.  The 1:1:1 configurations we investigated were composed of steel:bronze:PTFE (steel striker, 36 particles), steel:glass:nylon (steel striker, 33 particles), and steel:PTFE:rubber (steel striker, 27 particles).  Our specific choices of configurations were motivated by the desire to examine various combinations of hard, soft, heavy, and light materials (see the summary in Table~\ref{trimerprop}).

\begin{table}
\centerline{
\begin{tabular}{| l | l | l | l |} \hline
Configuration & Properties of Bead 1 & Properties of Bead 2 & Properties of Bead 3 \\ \hline 
Steel:Bronze:PTFE & large mass, large $E$, small $\nu$ & large mass, medium $E$, medium $\nu$ & small mass, small $E$, large $\nu$ \\ \hline
Steel:PTFE:Rubber & large mass, large $E$, small $\nu$ & small mass, small $E$, large $\nu$ & very small mass, very small $E$, large $\nu$ \\ \hline
Steel:Glass:Nylon & large mass, large $E$, small $\nu$ & small mass, medium $E$, very small $\nu$ & very small mass, small $E$, medium $\nu$ \\ \hline
\end{tabular}}
\caption{Summary of the trimer configurations that we investigated in terms of their relative masses, elastic moduli $E$, and Poisson ratios $\nu$.
}
\label{trimerprop}
\end{table}

\begin{figure}[tbp]
\centering \includegraphics[width=15.0cm]{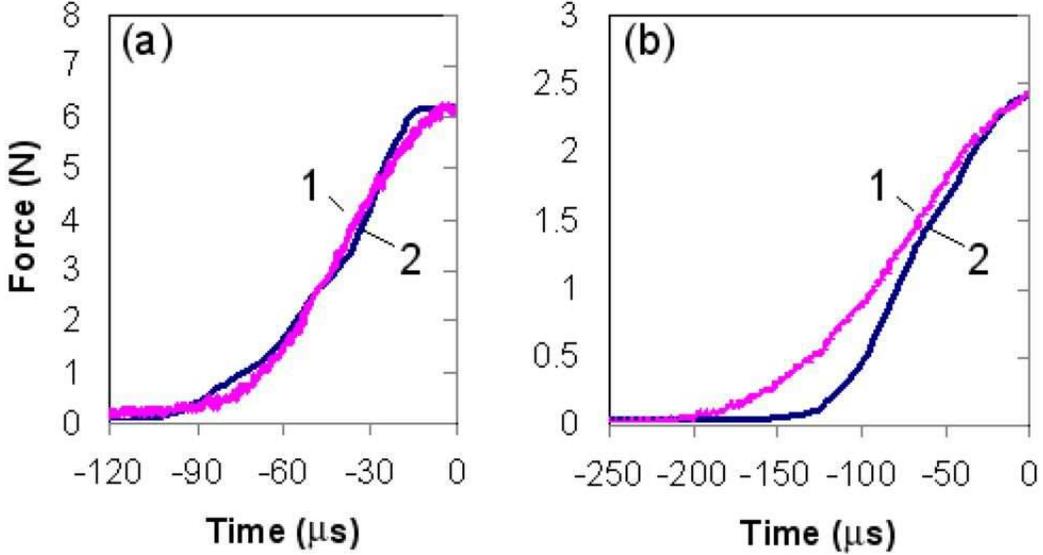}
\caption{(Color online) Rising shape comparison of two different solitary waves recorded in the middle of the chain for different trimer configurations.  Each plot displays the evolution of the force (in N) with time (in microseconds). Curve 1 corresponds to the experimental points, and curve 2 corresponds to the numerical ones. The depicted examples are (a) steel-glass-nylon trimer curves detected in particle 13 and (b) steel-bronze-PTFE curves detected in particle 12. Note that in order to compare the shapes, the experimental curves have been multiplied by an arbitrary factor to obtain the same force amplitude as that found in the numerical data (where no dissipation is present). 
}
\label{directtrimer}
\end{figure}

\begin{figure}[h]
\centerline{\includegraphics[height=7.5cm]{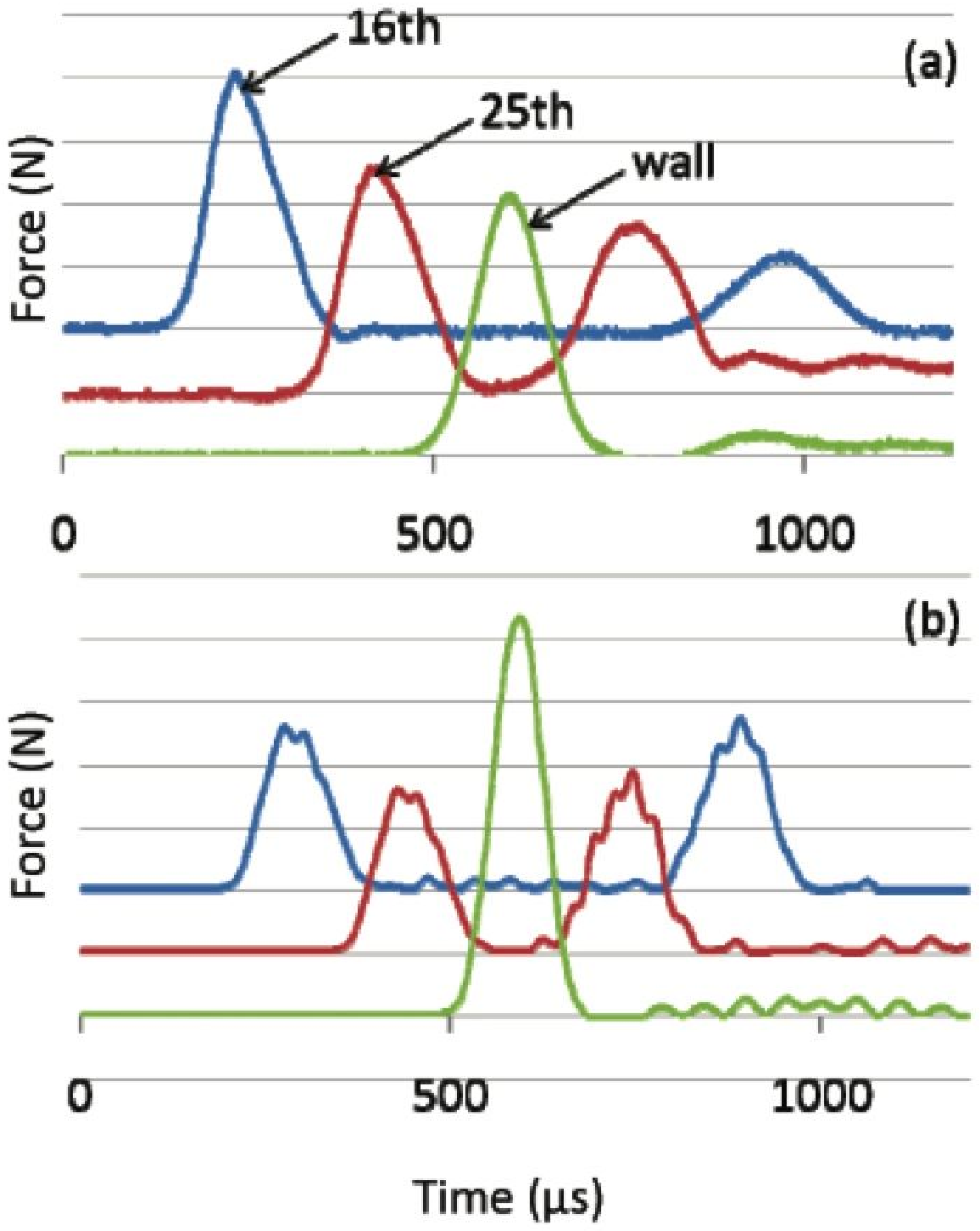}
\includegraphics[height=7.5cm]{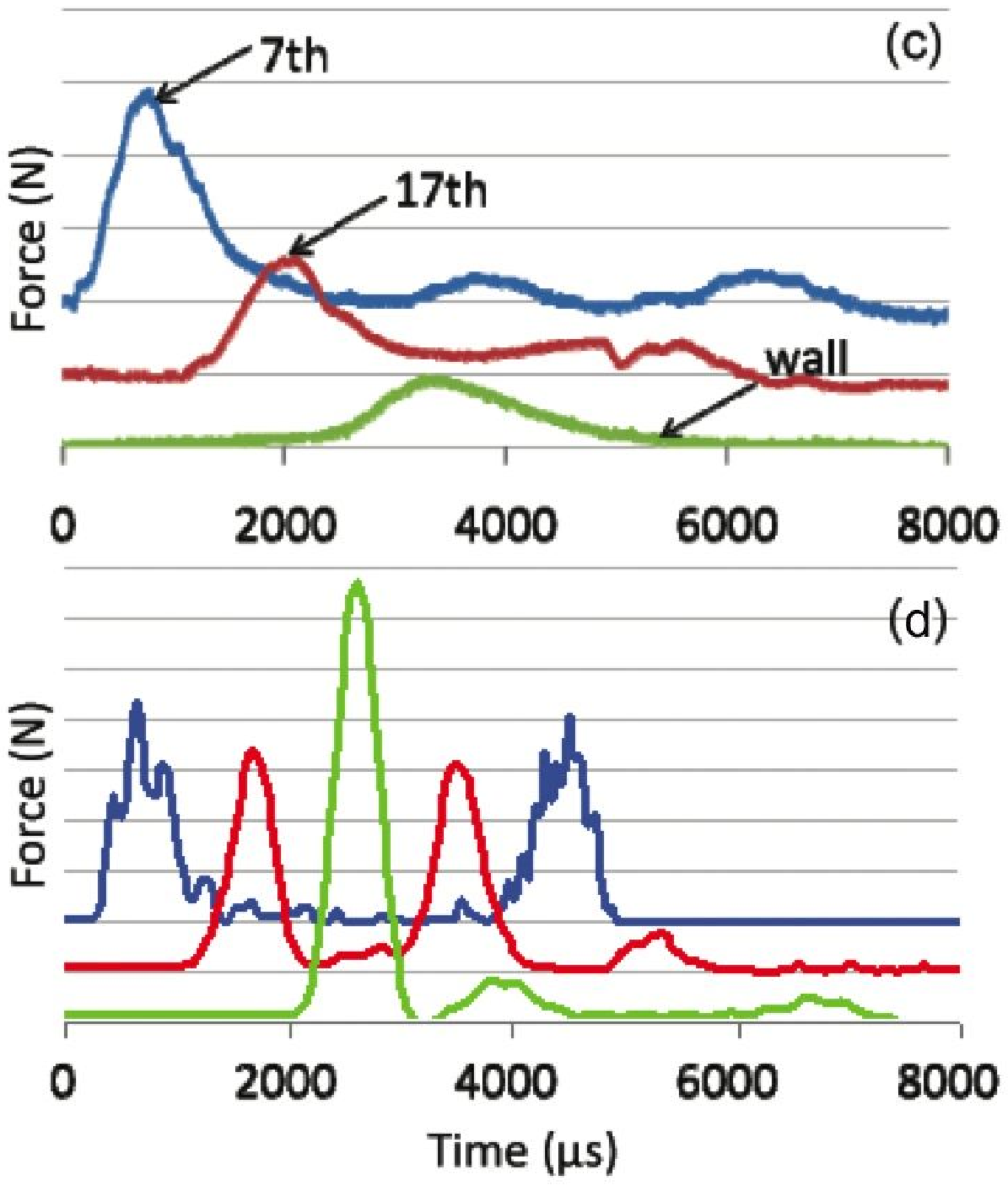}}
\caption{(Color online) Force versus time response for chains of 1:1:1 trimers consisting of (a,b) steel:glass:nylon and (c,d) steel:PTFE:rubber.  Panels (a,c) show experimental results and panels (b,d) show numerical results.  In (a,b), a steel striker was dropped with a velocity of 0.89 m/s on impact;
the $y$-axis scale is 1 N per division in (a) and 2 N per division in (b).  In (c,d), a steel striker was dropped with a velocity of 1.72 m/s on impact;
the $y$-axis scale is 0.5 N per division in both (c) and (d).  The numbered arrows point to the corresponding particles in the chain.  In panels (a,b), both depicted beads are made of steel; in panels (c,d), the 7th bead is made of steel and the 17th is made of PTFE. 
}
\label{othertrimers}
\end{figure}

As with the dimer chains, we calculated the pulse speed for each configuration using time-of-flight measurements between the sensors.  We obtained very good qualitative and reasonable quantitative agreement (very good, given the lack of dissipation in the models) between the results of experiments and numerics.  In Fig.~\ref{oneoneone}(b,c), we show the force versus time response for a chain of 1:1:1 steel:bronze:PTFE trimers.  We observed the formation of robust solitary waves with a pulse propagation speed between the first and second sensor of about 115 m/s experimentally and 160 m/s numerically.  The FWHM is about 4.2 particles (1.4 cells) experimentally and 4.3 particles (1.4 cells) numerically.  This 
illustrates the very good agreement between experiments and numerical simulations, as also indicated by the direct comparison of the pulses in Fig.~\ref{directtrimer}.  We also obtained very good agreement between experiments and numerics in the other  configurations.  In Fig.~\ref{othertrimers}, we show the force versus time response for chains of steel:glass:nylon trimers (left panels) and steel:PTFE:rubber trimers (right panels).  The steel:glass:nylon trimer chain had a steel striker dropped from 4 cm; its pulse speed was about 220 m/s experimentally and 286 m/s numerically.  The FWHM was about 1.6 cells 
both experimentally and numerically.  The steel:PTFE:rubber trimer chain had a steel striker dropped from 15 cm; its pulse velocity was about 33.9 m/s experimentally and 56.0 m/s numerically.  (Recall that the dissipative effects are larger for rubber than for other materials.)  The FWHM was about 1.4 cells both experimentally and numerically.  

For each configuration, we robustly demonstrated the formation and propagation of solitary waves.  These waves can be clearly discerned both experimentally and numerically for the steel:bronze:PTFE and Steel:PTFE:nylon chains but only numerically for the trimer chain containing rubber, in which dissipation is much more important.  The presence of small secondary pulses in the configurations can also typically be observed.  Additionally, one can see in both the experiments and the numerical simulations that the pulses are not perfectly symmetric near the top of the chains.  This skewness disappears as a pulse moves down a chain.

 \section{Conclusions}
 
We examined the propagation of solitary waves in heterogeneous, periodic granular media consisting of chains of beads using experiments, numerical simulations, and theory.  Specifically, we investigated 1D granular lattices of dimers and trimers composed of varying numbers of particles (yielding different periodicities) that were made of different materials.  We found both experimentally and numerically that such heterogeneous systems robustly support the formation and propagation of highly localized nonlinear solitary waves, with widths and pulse propagation speeds that depend on the periodicity of the chain (that is, the length of its unit cell).  

For dimer chains consisting of cells composed of $N_1$ particles of massive materials such as steel doped by 1 light particle such as PTFE, we find that the width (expressed as the number of unit cells) decreases, whereas the propagation speed increases with $N_1$.  We also observe a ``frustration" phenomenon for $N_1 \geq 4$, although robust localized pulses nevertheless form even in this case.  We showed numerically, experimentally, and analytically (using a long-wavelength approximation) that the force-velocity scaling for chains of dimers follows the same relation as that for homogeneous chains.  We also found very good agreement for the solitary-wave widths (which depend on the mass ratio of the dimer materials) using all three approaches, generalizing previous results from all known limiting cases.  Finally, we demonstrated both experimentally and numerically the formation and 
propagation of robust, strongly localized solitary waves in trimer lattices consisting of chains of cells composed of 1 particle of each of three varieties.  

The qualitative and quantitative understanding of the dynamics of dimer and trimer chains obtained in the present work, and their characterization as possessing highly nonlinear solitary waves with similarities (e.g., force-velocity scaling) as well as differences (e.g., in the width) from the standard monomer chain, paves the way for studies in increasingly heterogeneous media in both one and in higher dimensions. Such studies are currently in progress and will be reported in future publications.

 
\section*{Acknowledgements} 

M.A.P. acknowledges support from the Gordon and Betty Moore Foundation through Caltech's Center for the Physics of Information (where he was a postdoc during much of this research) and the Graduate Aeronautical Laboratories at Caltech (GALCIT) for hosting one of his visits.  C.D. acknowledges support from start-up funds at Caltech, and P.G.K. acknowledges support from NSF-DMS and CAREER.  We thank Vitali F. Nesterenko for useful discussions.

\vspace{.1 in}

$^*$ Electronic address: {\it daraio@caltech.edu}


\begin{thebibliography}{46}
\expandafter\ifx\csname natexlab\endcsname\relax\def\natexlab#1{#1}\fi
\expandafter\ifx\csname bibnamefont\endcsname\relax
  \def\bibnamefont#1{#1}\fi
\expandafter\ifx\csname bibfnamefont\endcsname\relax
  \def\bibfnamefont#1{#1}\fi
\expandafter\ifx\csname citenamefont\endcsname\relax
  \def\citenamefont#1{#1}\fi
\expandafter\ifx\csname url\endcsname\relax
  \def\url#1{\texttt{#1}}\fi
\expandafter\ifx\csname urlprefix\endcsname\relax\def\urlprefix{URL }\fi
\providecommand{\bibinfo}[2]{#2}
\providecommand{\eprint}[2][]{\url{#2}}

\bibitem[{\citenamefont{Fermi et~al.}(1955)\citenamefont{Fermi, Pasta, and
  Ulam}}]{fpu55}
\bibinfo{author}{\bibfnamefont{E.}~\bibnamefont{Fermi}},
  \bibinfo{author}{\bibfnamefont{J.~R.} \bibnamefont{Pasta}}, \bibnamefont{and}
  \bibinfo{author}{\bibfnamefont{S.}~\bibnamefont{Ulam}}, \bibinfo{type}{Tech.
  Rep.} \bibinfo{number}{Report LA-1940}, \bibinfo{institution}{Los Alamos}
  (\bibinfo{year}{1955}).

\bibitem[{\citenamefont{Campbell et~al.}(2005)\citenamefont{Campbell, Rosenau,
  and Zaslavsky}}]{focus}
\bibinfo{author}{\bibfnamefont{D.~K.} \bibnamefont{Campbell}},
  \bibinfo{author}{\bibfnamefont{P.}~\bibnamefont{Rosenau}}, \bibnamefont{and}
  \bibinfo{author}{\bibfnamefont{G.}~\bibnamefont{Zaslavsky}},
  \bibinfo{journal}{Chaos} \textbf{\bibinfo{volume}{15}},
  \bibinfo{pages}{015101} (\bibinfo{year}{2005}).

\bibitem[{\citenamefont{Ford}(1992)}]{Ford}
\bibinfo{author}{\bibfnamefont{J.}~\bibnamefont{Ford}},
  \bibinfo{journal}{Phys. Rep.} \textbf{\bibinfo{volume}{213}},
  \bibinfo{pages}{271} (\bibinfo{year}{1992}).

\bibitem[{\citenamefont{Lichtenberg and Lieberman}(1992)}]{lich}
\bibinfo{author}{\bibfnamefont{A.~J.} \bibnamefont{Lichtenberg}}
  \bibnamefont{and} \bibinfo{author}{\bibfnamefont{M.~A.}
  \bibnamefont{Lieberman}}, \emph{\bibinfo{title}{Regular and Chaotic
  Dynamics}}, no.~\bibinfo{number}{38} in \bibinfo{series}{Applied Mathematical
  Sciences} (\bibinfo{publisher}{Springer-Verlag}, \bibinfo{address}{New York,
  NY}, \bibinfo{year}{1992}), \bibinfo{edition}{2nd} ed.

\bibitem[{\citenamefont{Morsch and Oberthaler}(2006)}]{morsch1}
\bibinfo{author}{\bibfnamefont{O.}~\bibnamefont{Morsch}} \bibnamefont{and}
  \bibinfo{author}{\bibfnamefont{M.}~\bibnamefont{Oberthaler}},
  \bibinfo{journal}{Rev. Mod. Phys.} \textbf{\bibinfo{volume}{78}},
  \bibinfo{pages}{179} (\bibinfo{year}{2006}).

\bibitem[{\citenamefont{Brazhnyi and Konotop}(2004)}]{konotopmplb}
\bibinfo{author}{\bibfnamefont{V.~A.} \bibnamefont{Brazhnyi}} \bibnamefont{and}
  \bibinfo{author}{\bibfnamefont{V.~V.} \bibnamefont{Konotop}},
  \bibinfo{journal}{Mod. Phys. Lett. B} \textbf{\bibinfo{volume}{18}},
  \bibinfo{pages}{627} (\bibinfo{year}{2004}).

\bibitem[{\citenamefont{Porter et~al.}(2005)\citenamefont{Porter,
  Carretero-Gonz\'alez, Kevrekidis, and Malomed}}]{fpubec}
\bibinfo{author}{\bibfnamefont{M.~A.} \bibnamefont{Porter}},
  \bibinfo{author}{\bibfnamefont{R.}~\bibnamefont{Carretero-Gonz\'alez}},
  \bibinfo{author}{\bibfnamefont{P.~G.} \bibnamefont{Kevrekidis}},
  \bibnamefont{and} \bibinfo{author}{\bibfnamefont{B.~A.}
  \bibnamefont{Malomed}}, \bibinfo{journal}{Chaos}
  \textbf{\bibinfo{volume}{15}}, \bibinfo{pages}{015115} (\bibinfo{year}{2005}).

\bibitem[{\citenamefont{Kivshar and Agrawal}(2003)}]{photon}
\bibinfo{author}{\bibfnamefont{Y.~S.} \bibnamefont{Kivshar}} \bibnamefont{and}
  \bibinfo{author}{\bibfnamefont{G.~P.} \bibnamefont{Agrawal}},
  \emph{\bibinfo{title}{Optical Solitons: From Fibers to Photonic Crystals}}
  (\bibinfo{publisher}{Academic Press}, \bibinfo{address}{San Diego,
  California}, \bibinfo{year}{2003}).

\bibitem[{\citenamefont{Fleischer et~al.}(2005)\citenamefont{Fleischer, Bartal,
  Cohen, Schwartz, Manela, Freedman, Segev, Buljan, and Efremidis}}]{efrem1}
\bibinfo{author}{\bibfnamefont{J.}~\bibnamefont{Fleischer}},
  \bibinfo{author}{\bibfnamefont{G.}~\bibnamefont{Bartal}},
  \bibinfo{author}{\bibfnamefont{O.}~\bibnamefont{Cohen}},
  \bibinfo{author}{\bibfnamefont{T.}~\bibnamefont{Schwartz}},
  \bibinfo{author}{\bibfnamefont{O.}~\bibnamefont{Manela}},
  \bibinfo{author}{\bibfnamefont{B.}~\bibnamefont{Freedman}},
  \bibinfo{author}{\bibfnamefont{M.}~\bibnamefont{Segev}},
  \bibinfo{author}{\bibfnamefont{H.}~\bibnamefont{Buljan}}, \bibnamefont{and}
  \bibinfo{author}{\bibfnamefont{N.}~\bibnamefont{Efremidis}},
  \bibinfo{journal}{Opt. Expr.} \textbf{\bibinfo{volume}{13}},
  \bibinfo{pages}{1780} (\bibinfo{year}{2005}).

\bibitem[{\citenamefont{Peyrard}(2004)}]{peyrard}
\bibinfo{author}{\bibfnamefont{M.}~\bibnamefont{Peyrard}},
  \bibinfo{journal}{Nonlinearity} \textbf{\bibinfo{volume}{17}},
  \bibinfo{pages}{R1} (\bibinfo{year}{2004}).

\bibitem[{\citenamefont{Kivshar and Flach}(2003)}]{focus2}
\bibinfo{author}{\bibfnamefont{Y.~S.} \bibnamefont{Kivshar}} \bibnamefont{and}
  \bibinfo{author}{\bibfnamefont{S.}~\bibnamefont{Flach}},
  \bibinfo{journal}{Chaos} \textbf{\bibinfo{volume}{13}}, \bibinfo{pages}{586}
  (\bibinfo{year}{2003}).

\bibitem[{\citenamefont{Nesterenko}(2001)}]{nesterenko1}
\bibinfo{author}{\bibfnamefont{V.~F.} \bibnamefont{Nesterenko}},
  \emph{\bibinfo{title}{Dynamics of Heterogeneous Materials}}
  (\bibinfo{publisher}{Springer-Verlag}, \bibinfo{address}{New York, NY},
  \bibinfo{year}{2001}).

\bibitem[{\citenamefont{Coste et~al.}(1997)\citenamefont{Coste, Falcon, and
  Fauve}}]{coste97}
\bibinfo{author}{\bibfnamefont{C.}~\bibnamefont{Coste}},
  \bibinfo{author}{\bibfnamefont{E.}~\bibnamefont{Falcon}}, \bibnamefont{and}
  \bibinfo{author}{\bibfnamefont{S.}~\bibnamefont{Fauve}},
  \bibinfo{journal}{Phys. Rev. E} \textbf{\bibinfo{volume}{56}},
  \bibinfo{pages}{6104} (\bibinfo{year}{1997}).

\bibitem[{\citenamefont{Daraio et~al.}(2006{\natexlab{a}})\citenamefont{Daraio,
  Nesterenko, Herbold, and Jin}}]{nesterenko2}
\bibinfo{author}{\bibfnamefont{C.}~\bibnamefont{Daraio}},
  \bibinfo{author}{\bibfnamefont{V.~F.} \bibnamefont{Nesterenko}},
  \bibinfo{author}{\bibfnamefont{E.~B.} \bibnamefont{Herbold}},
  \bibnamefont{and} \bibinfo{author}{\bibfnamefont{S.}~\bibnamefont{Jin}},
  \bibinfo{journal}{Phys. Rev. E} \textbf{\bibinfo{volume}{73}}, \bibinfo{eid}{026610} (\bibinfo{year}{2006}{\natexlab{a}})

\bibitem[{\citenamefont{Daraio et~al.}(2005)\citenamefont{Daraio, Nesterenko,
  Herbold, and Jin}}]{dar05}
\bibinfo{author}{\bibfnamefont{C.}~\bibnamefont{Daraio}},
  \bibinfo{author}{\bibfnamefont{V.~F.} \bibnamefont{Nesterenko}},
  \bibinfo{author}{\bibfnamefont{E.~B.} \bibnamefont{Herbold}},
  \bibnamefont{and} \bibinfo{author}{\bibfnamefont{S.}~\bibnamefont{Jin}},
  \bibinfo{journal}{Phys. Rev. E} \textbf{\bibinfo{volume}{72}},
  \bibinfo{pages}{016603} (\bibinfo{year}{2005}).

\bibitem[{\citenamefont{Nesterenko et~al.}(2005)\citenamefont{Nesterenko,
  Daraio, Herbold, and Jin}}]{dar05b}
\bibinfo{author}{\bibfnamefont{V.~F.} \bibnamefont{Nesterenko}},
  \bibinfo{author}{\bibfnamefont{C.}~\bibnamefont{Daraio}},
  \bibinfo{author}{\bibfnamefont{E.~B.} \bibnamefont{Herbold}},
  \bibnamefont{and} \bibinfo{author}{\bibfnamefont{S.}~\bibnamefont{Jin}},
  \bibinfo{journal}{Phys. Rev. Lett.} \textbf{\bibinfo{volume}{95}},
  \bibinfo{eid}{158702} (\bibinfo{year}{2005}).

\bibitem[{\citenamefont{Daraio et~al.}(2006{\natexlab{b}})\citenamefont{Daraio,
  Nesterenko, Herbold, and Jin}}]{dar06}
\bibinfo{author}{\bibfnamefont{C.}~\bibnamefont{Daraio}},
  \bibinfo{author}{\bibfnamefont{V.~F.} \bibnamefont{Nesterenko}},
  \bibinfo{author}{\bibfnamefont{E.~B.} \bibnamefont{Herbold}},
  \bibnamefont{and} \bibinfo{author}{\bibfnamefont{S.}~\bibnamefont{Jin}},
  \bibinfo{journal}{Phys. Rev. Lett.} \textbf{\bibinfo{volume}{96}},
  \bibinfo{eid}{058002} (\bibinfo{year}{2006}{\natexlab{b}}).

\bibitem[{\citenamefont{Hasco\"{e}t and Herrmann}(2000)}]{hascoet00}
\bibinfo{author}{\bibfnamefont{E.}~\bibnamefont{Hasco\"{e}t}} \bibnamefont{and}
  \bibinfo{author}{\bibfnamefont{H.~J.} \bibnamefont{Herrmann}},
  \bibinfo{journal}{Euro. Phys. Journal B} \textbf{\bibinfo{volume}{14}}, \bibinfo{pages}{183}
  (\bibinfo{year}{2000}).

\bibitem[{\citenamefont{Hinch and Saint-Jean}(1999)}]{hinch99}
\bibinfo{author}{\bibfnamefont{E.~J.} \bibnamefont{Hinch}} \bibnamefont{and}
  \bibinfo{author}{\bibfnamefont{S.}~\bibnamefont{Saint-Jean}},
  \bibinfo{journal}{Proc. Royal Soc. London A}
  \textbf{\bibinfo{volume}{455}}, \bibinfo{pages}{3201} (\bibinfo{year}{1999}).

\bibitem[{\citenamefont{Manciu et~al.}(2001)\citenamefont{Manciu, Sen, and
  Hurd}}]{man01}
\bibinfo{author}{\bibfnamefont{M.}~\bibnamefont{Manciu}},
  \bibinfo{author}{\bibfnamefont{S.}~\bibnamefont{Sen}}, \bibnamefont{and}
  \bibinfo{author}{\bibfnamefont{A.~J.} \bibnamefont{Hurd}},
  \bibinfo{journal}{Physica D} \textbf{\bibinfo{volume}{157}},
  \bibinfo{pages}{226} (\bibinfo{year}{2001}).

\bibitem[{\citenamefont{Hong and Xu}(2002)}]{hong02}
\bibinfo{author}{\bibfnamefont{J.}~\bibnamefont{Hong}} \bibnamefont{and}
  \bibinfo{author}{\bibfnamefont{A.}~\bibnamefont{Xu}},
  \bibinfo{journal}{App. Phys. Lett.} \textbf{\bibinfo{volume}{81}},
  \bibinfo{pages}{4868} (\bibinfo{year}{2002}).

\bibitem[{\citenamefont{Hong}(2005)}]{hong05}
\bibinfo{author}{\bibfnamefont{J.}~\bibnamefont{Hong}},
  \bibinfo{journal}{Phys. Rev. Lett.} \textbf{\bibinfo{volume}{94}},
  \bibinfo{eid}{108001} (\bibinfo{year}{2005}).

\bibitem[{\citenamefont{Job et~al.}(2005)\citenamefont{Job, Melo, Sokolow, and
  Sen}}]{job05}
\bibinfo{author}{\bibfnamefont{S.}~\bibnamefont{Job}},
  \bibinfo{author}{\bibfnamefont{F.}~\bibnamefont{Melo}},
  \bibinfo{author}{\bibfnamefont{A.}~\bibnamefont{Sokolow}}, \bibnamefont{and}
  \bibinfo{author}{\bibfnamefont{S.}~\bibnamefont{Sen}},
  \bibinfo{journal}{Phys. Rev. Lett.} \textbf{\bibinfo{volume}{94}},
  \bibinfo{eid}{178002} (\bibinfo{year}{2005}).

\bibitem[{\citenamefont{Doney and Sen}(2006)}]{doney06}
\bibinfo{author}{\bibfnamefont{R.}~\bibnamefont{Doney}} \bibnamefont{and}
  \bibinfo{author}{\bibfnamefont{S.}~\bibnamefont{Sen}},
  \bibinfo{journal}{Phys. Rev. Lett.} \textbf{\bibinfo{volume}{97}},
  \bibinfo{eid}{155502} (\bibinfo{year}{2006}).

\bibitem[{\citenamefont{Vergara}(2006)}]{ver06}
\bibinfo{author}{\bibfnamefont{L.}~\bibnamefont{Vergara}},
  \bibinfo{journal}{Phys. Rev. E} \textbf{\bibinfo{volume}{73}}, \bibinfo{eid}{066623}
  (\bibinfo{year}{2006}).

\bibitem[{\citenamefont{Sokolow et~al.}(2007)\citenamefont{Sokolow, Bittle, and
  Sen}}]{sok07}
\bibinfo{author}{\bibfnamefont{A.}~\bibnamefont{Sokolow}},
  \bibinfo{author}{\bibfnamefont{E.~G.} \bibnamefont{Bittle}},
  \bibnamefont{and} \bibinfo{author}{\bibfnamefont{S.}~\bibnamefont{Sen}},
  \bibinfo{journal}{Europhys. Lett.} \textbf{\bibinfo{volume}{77}}, \bibinfo{eid}{24002}
  (\bibinfo{year}{2007}).

\bibitem[{\citenamefont{Herbold et~al.}(2006)\citenamefont{Herbold, Nesterenko,
  and Daraio}}]{herb06}
\bibinfo{author}{\bibfnamefont{E.~B.} \bibnamefont{Herbold}},
  \bibinfo{author}{\bibfnamefont{V.~F.} \bibnamefont{Nesterenko}},
  \bibnamefont{and} \bibinfo{author}{\bibfnamefont{C.}~\bibnamefont{Daraio}},
  in \emph{\bibinfo{booktitle}{APS - Shock Compression of Condensed Matter}}
  (\bibinfo{publisher}{AIP Conference Proceedings},
  \bibinfo{address}{Baltimore, MD}, \bibinfo{year}{2006}), pp.
  \bibinfo{pages}{1523--1526}.

\bibitem[{\citenamefont{Hong and Xu}(2001)}]{hong01b}
\bibinfo{author}{\bibfnamefont{J.}~\bibnamefont{Hong}} \bibnamefont{and}
  \bibinfo{author}{\bibfnamefont{A.}~\bibnamefont{Xu}},
  \bibinfo{journal}{Phys. Rev. E} \textbf{\bibinfo{volume}{63}},
  \bibinfo{pages}{061310} (\bibinfo{year}{2001}).

\bibitem[{\citenamefont{Rosenau and Hyman}(1993)}]{rosenau1}
\bibinfo{author}{\bibfnamefont{P.}~\bibnamefont{Rosenau}} \bibnamefont{and}
  \bibinfo{author}{\bibfnamefont{J.~M.} \bibnamefont{Hyman}},
  \bibinfo{journal}{Phys. Rev. Lett.} \textbf{\bibinfo{volume}{70}},
  \bibinfo{pages}{564} (\bibinfo{year}{1993}).

\bibitem[{\citenamefont{Porter et~al.}(In press)\citenamefont{Porter, Daraio,
  Herbold, Szelengowicz, and Kevrekidis}}]{dimer}
\bibinfo{author}{\bibfnamefont{M.~A.} \bibnamefont{Porter}},
  \bibinfo{author}{\bibfnamefont{C.}~\bibnamefont{Daraio}},
  \bibinfo{author}{\bibfnamefont{E.~B.} \bibnamefont{Herbold}},
  \bibinfo{author}{\bibfnamefont{I.}~\bibnamefont{Szelengowicz}},
  \bibnamefont{and} \bibinfo{author}{\bibfnamefont{P.~G.}
  \bibnamefont{Kevrekidis}}, \bibinfo{journal}{Phys. Rev. E (Rapid
  Comm.)}  (\bibinfo{year}{In press}). 

\bibitem[{\citenamefont{Dash and Patnaik}(1981)}]{dimer81}
\bibinfo{author}{\bibfnamefont{P.~C.} \bibnamefont{Dash}} \bibnamefont{and}
  \bibinfo{author}{\bibfnamefont{K.}~\bibnamefont{Patnaik}},
  \bibinfo{journal}{Prog. Theor. Phys.}
  \textbf{\bibinfo{volume}{65}}, \bibinfo{pages}{1526} (\bibinfo{year}{1981}).

\bibitem[{\citenamefont{Bilz et~al.}(1982)\citenamefont{Bilz, B\"uttner,
  Bussmann-Holder, Kress, and Schr\"oder}}]{ferro}
\bibinfo{author}{\bibfnamefont{H.}~\bibnamefont{Bilz}},
  \bibinfo{author}{\bibfnamefont{H.}~\bibnamefont{B\"uttner}},
  \bibinfo{author}{\bibfnamefont{A.}~\bibnamefont{Bussmann-Holder}},
  \bibinfo{author}{\bibfnamefont{W.}~\bibnamefont{Kress}}, \bibnamefont{and}
  \bibinfo{author}{\bibfnamefont{U.}~\bibnamefont{Schr\"oder}},
  \bibinfo{journal}{Phys. Rev. Lett.} \textbf{\bibinfo{volume}{48}},
  \bibinfo{pages}{264} (\bibinfo{year}{1982}).

\bibitem[{\citenamefont{Rice and Mele}(1982)}]{polymer}
\bibinfo{author}{\bibfnamefont{M.~J.} \bibnamefont{Rice}} \bibnamefont{and}
  \bibinfo{author}{\bibfnamefont{E.~J.} \bibnamefont{Mele}},
  \bibinfo{journal}{Phys. Rev. Lett.} \textbf{\bibinfo{volume}{49}},
  \bibinfo{pages}{1455} (\bibinfo{year}{1982}).

\bibitem[{\citenamefont{Sukhorukov and Kivshar}(2002)}]{sukh02}
\bibinfo{author}{\bibfnamefont{A.~A.} \bibnamefont{Sukhorukov}}
  \bibnamefont{and} \bibinfo{author}{\bibfnamefont{Y.~S.}
  \bibnamefont{Kivshar}}, \bibinfo{journal}{Opt. Lett.}
  \textbf{\bibinfo{volume}{27}}, \bibinfo{pages}{2112} (\bibinfo{year}{2002}).

\bibitem[{\citenamefont{Morandotti et~al.}(2002)\citenamefont{Morandotti,
  Mandelik, Silberberg, Aitchison, Sorel, Christodoulides, Sukhorukov, and
  Kivshar}}]{moran04}
\bibinfo{author}{\bibfnamefont{R.}~\bibnamefont{Morandotti}},
  \bibinfo{author}{\bibfnamefont{D.}~\bibnamefont{Mandelik}},
  \bibinfo{author}{\bibfnamefont{Y.}~\bibnamefont{Silberberg}},
  \bibinfo{author}{\bibfnamefont{J.~S.} \bibnamefont{Aitchison}},
  \bibinfo{author}{\bibfnamefont{M.}~\bibnamefont{Sorel}},
  \bibinfo{author}{\bibfnamefont{D.~N.} \bibnamefont{Christodoulides}},
  \bibinfo{author}{\bibfnamefont{A.~A.} \bibnamefont{Sukhorukov}},
  \bibnamefont{and} \bibinfo{author}{\bibfnamefont{Y.~S.}
  \bibnamefont{Kivshar}}, \bibinfo{journal}{Opt. Lett.}
  \textbf{\bibinfo{volume}{29}}, \bibinfo{pages}{2890} (\bibinfo{year}{2002}).

\bibitem[{\citenamefont{Sato et~al.}(2006)\citenamefont{Sato, Hubbard, and
  Sievers}}]{sievers}
\bibinfo{author}{\bibfnamefont{M.}~\bibnamefont{Sato}},
  \bibinfo{author}{\bibfnamefont{B.~E.} \bibnamefont{Hubbard}},
  \bibnamefont{and} \bibinfo{author}{\bibfnamefont{A.~J.}
  \bibnamefont{Sievers}}, \bibinfo{journal}{Rev. of Mod. Phys.}
  \textbf{\bibinfo{volume}{78}}, \bibinfo{pages}{137} (\bibinfo{year}{2006}).

\bibitem[{met(1983)}]{metals}
\emph{\bibinfo{title}{ASM Metals Reference Book}} (\bibinfo{publisher}{American
  Society for Metals}, \bibinfo{address}{Metals Park, OH},
  \bibinfo{year}{1983}), \bibinfo{edition}{2nd} ed.

\bibitem[{316()}]{316}
\bibinfo{note}{\url{http://www.efunda.com}}.

\bibitem[{dup()}]{dupont}
\bibinfo{note}{\url{www.dupont.com/teflon/chemical/}}.

\bibitem[{\citenamefont{Carter and Marsh}(1955)}]{carter95}
\bibinfo{author}{\bibfnamefont{W.~J.} \bibnamefont{Carter}} \bibnamefont{and}
  \bibinfo{author}{\bibfnamefont{S.~P.} \bibnamefont{Marsh}},
  \bibinfo{type}{Tech. Rep.} \bibinfo{number}{Report No. LA-13006-MS},
  \bibinfo{institution}{Los Alamos} (\bibinfo{year}{1955}).

\bibitem[{bro()}]{bronze}
\bibinfo{note}{\url{http://www.matweb.com}}.

\bibitem[{\citenamefont{Benton}(1901)}]{bronzepoisson}
\bibinfo{author}{\bibfnamefont{J.~R.} \bibnamefont{Benton}},
  \bibinfo{journal}{Phys. Rev. (Series I)} \textbf{\bibinfo{volume}{12}},
  \bibinfo{pages}{36} (\bibinfo{year}{1901}).

\bibitem[{gla()}]{glass}
\bibinfo{note}{\url{http://www.camglassblowing.co.uk/gproperties.htm}}.

\bibitem[{\citenamefont{Coste and Gilles}(1999)}]{coste99}
\bibinfo{author}{\bibfnamefont{C.}~\bibnamefont{Coste}} \bibnamefont{and}
  \bibinfo{author}{\bibfnamefont{B.}~\bibnamefont{Gilles}},
  \bibinfo{journal}{Euro. Phys. Journal B}
  \textbf{\bibinfo{volume}{7}}, \bibinfo{pages}{155} (\bibinfo{year}{1999}).

\bibitem[{\citenamefont{Peyrard and Kruskal}(1984)}]{peykru}
\bibinfo{author}{\bibfnamefont{M.}~\bibnamefont{Peyrard}} \bibnamefont{and}
  \bibinfo{author}{\bibfnamefont{M.}~\bibnamefont{Kruskal}},
  \bibinfo{journal}{Physica D} \textbf{\bibinfo{volume}{14}},
  \bibinfo{pages}{88} (\bibinfo{year}{1984}).

\bibitem[{\citenamefont{Pnevmatikos et~al.}(1986)\citenamefont{Pnevmatikos,
  Flytzanis, and Remoissenet}}]{pnevmatikos}
\bibinfo{author}{\bibfnamefont{S.}~\bibnamefont{Pnevmatikos}},
  \bibinfo{author}{\bibfnamefont{N.}~\bibnamefont{Flytzanis}},
  \bibnamefont{and}
  \bibinfo{author}{\bibfnamefont{M.}~\bibnamefont{Remoissenet}},
  \bibinfo{journal}{Phys. Rev. B} \textbf{\bibinfo{volume}{33}},
  \bibinfo{pages}{2308} (\bibinfo{year}{1986}).

\end{thebibliography}

\end{document}